\documentclass[11pt, a4paper]{article}
\pdfoutput=1
\usepackage[dvipsnames]{xcolor}
\usepackage[utf8]{inputenc}
\usepackage{url}
\usepackage{amsmath}
\usepackage{amsfonts}
\usepackage{amssymb} 
\usepackage{graphicx}
\usepackage{array}
\usepackage{lipsum}
\usepackage{float}
\usepackage{multirow}
\usepackage{hhline}
\usepackage{tabularx}
\usepackage{subfigure}
\usepackage{afterpage}
\usepackage{epstopdf}
\usepackage[utf8]{inputenc}

\usepackage[pdftitle={(Quadratically) Refined Discrete Anomaly Cancellation},
  pdfauthor={Markus Dierigl, Michelangelo Tartaglia},
  pdfsubject={},
  bookmarksopen, bookmarksnumbered, bookmarksopenlevel=2, colorlinks=false, linkcolor=blue, citecolor=blue, urlcolor=blue]{hyperref}
\usepackage{comment}
\usepackage{rotating}
\usepackage{setspace}
\usepackage{longtable}
\usepackage{pdflscape}
\usepackage{caption}

\usepackage{array}
\newcolumntype{C}[1]{>{\centering\let\newline\\\arraybackslash\hspace{0pt}}m{#1}}

\usepackage{tikz}
\usepackage{tikz-3dplot}
\usetikzlibrary{calc}
\usetikzlibrary{decorations} 
\usepackage{tikz-cd} 

\usepackage[all]{xy}

\numberwithin{equation}{section}

\usepackage[nosort]{cite}

\newcommand{\QR}{\mathcal{Q}}

\newcommand{\Z}{\mathbb{Z}}
\newcommand{\R}{\mathbb{R}}

\begin{document}

\baselineskip=18pt

\vspace*{-2cm}
\begin{flushright}
  \texttt{CERN-TH-2025-068} \\

  \texttt{IFT-25-35}
\end{flushright}

\vspace*{0.6cm} 
\begin{center}
{\Large{\textbf{(Quadratically) Refined Discrete Anomaly Cancellation}}} \\
 \vspace*{1.5cm}
Markus Dierigl$^1$, Michelangelo Tartaglia$^{2}$\\

{
 \vspace*{1.0cm} 
{\it 
${}^1$ Theoretical Physics Department, CERN, 1211 Geneva 23, Switzerland\\ 
${}^2$Instituto de F\'{i}sica Te\'{o}rica IFT-UAM/CSIC,
C/ Nicol\'{a}s Cabrera 13-15, Campus de Cantoblanco, 28049 Madrid, Spain 
}}

\vspace*{0.8cm}
\end{center}
\vspace*{.5cm}

\noindent In this work we study the cancellation of non-perturbative anomalies of gravitational theories with gauge group $\mathbb{Z}_k$ in six dimensions. These subtle anomalies require a classification of deformation classes of manifolds with discrete gauge bundles known as bordism groups. The consistency of the theory demands a cancellation of  the fermion anomalies, which can be done by the transformation properties of 2-form fields in the theory. Since the 2-forms in six dimensions are themselves chiral, their formulation needs subtle topological information encoded in a so-called quadratic refinement. A matching between the fermionic anomalies and the defining properties of the quadratic refinement, lead to strong consistency constraints on the charged fermion spectrum. We explicitly determine these consistency conditions for the case of a single chiral 2-form and various discrete gauge groups. Since we provide a model-independent formulation, these restrictions hold universally for theories of this type.

\thispagestyle{empty}
\clearpage

\setcounter{page}{1}

\newpage
\tableofcontents

\newpage

\renewcommand{\arraystretch}{1.2}

\section{Introduction}
\label{sec:intro}

Gauge symmetries cannot be broken explicitly without making the underlying theory inconsistent. This is particularly subtle in case the breaking effects only occur at the quantum level via anomalies; see, e.g. \cite{Harvey:2005it, TachikawaLectures}. These anomalies are typically associated with the presence of chiral degrees of freedom such as Weyl fermions charged under the symmetry. One way to ensure the absence of anomalies is to modify the transformation behavior of other fields in the theory under gauge transformations. In particular, for the modification of a 2-form field in ten dimensions this form of anomaly cancellation is known as Green-Schwarz mechanism and ensures the gauge invariance of the heterotic string \cite{Green:1984sg}. A generalization of this can be used in six dimensions and was discussed for example in \cite{Green:1984bx, Sagnotti:1992qw}, canceling a subgroup of all possible gauge and gravitational anomalies.\footnote{For a more modern view on the Green-Schwarz mechanism see \cite{Lee:2022spd, Yonekura:2022reu, Tachikawa:2023lwf, Basile:2023zng, Kaidi:2024cbx}.}

In most of its applications, these generalized versions of the Green-Schwarz mechanism are used to cancel perturbative anomalies, i.e., anomalies that can be detected by infinitesimal gauge transformations or diffeomorphisms. These are canceled by modifying the transformation properties of a 2-form tensor field $B$, which now shifts under the anomalous transformations. In six-dimensional theories these considerations are particularly rich, since one has various non-trivial anomalies including the existence of pure gravitational effects. At the same time the 2-form fields in six dimensions are special, since they are generically chiral themselves. They consequently contribute to anomalies and can be used to cancel them. However, this chirality encoded in the the (anti-)self-duality of the tensor fields leads to various subtleties in their quantization, see, e.g., \cite{Witten:1996hc, Moore:1999gb, Freed:2000ta, Hopkins:2002rd, Belov:2006jd, Monnier:2016jlo, Monnier:2017klz, Monnier:2018nfs, Hsieh:2020jpj}.

In this work, we describe an anomaly cancellation mechanism based on (anti-)self-dual 2-form fields in six dimensions for the non-perturbative anomalies in the presence of a discrete gauge theory with gauge group $\mathbb{Z}_k$. Since the gauge symmetry is discrete there are no infinitesimal gauge transformations and in the absence of pure gravitational anomalies all anomalies are non-perturbative. Therefore, these theories are the ideal laboratories to explore generalized Green-Schwarz anomaly cancellation at the global level. The subtlety of the discrete gauge anomalies as well as the chiral nature of the tensor fields require introduction of an invertible anomaly theory in one higher dimension, developed in \cite{Dai:1994kq, Freed:2014iua, Witten:2015aba, Yonekura:2016wuc, Witten:2019bou, Hsieh:2020jpj} and applied in the context of discrete gauge theories, for example in \cite{Garcia-Etxebarria:2018ajm, Hsieh:2018ifc, Wan:2018bns, Davighi:2020uab, Davighi:2022icj, Dierigl:2022zll}. 

The description of chiral tensors depends on extra information encoded in a so-called quadratic refinement \cite{Hopkins:2002rd, Monnier:2018nfs, Hsieh:2020jpj}, which satisfies certain defining properties. This quadratic refinement also enters in the anomaly cancellation and influences which discrete fermion anomalies can be canceled, see, e.g., \cite{Dierigl:2022zll}. This allows us to use the general properties of quadratic refinements in order to heavily constrain the allowed spectrum of fermions charged under the discrete gauge group, ruling out many seemingly consistent theories. This form of anomaly cancellation further links the six-dimensional fermion spectrum to the chiral spectrum on the worldvolume of strings coupling to the 2-form fields under anomaly inflow. Therefore, our investigation demonstrates the effectiveness of the Green Schwarz mechanism and its extension to discrete symmetries but also points out its limitations. 

The general strategy is as follows: a successful anomaly cancellation equates the non-perturbative fermion anomaly and the Green-Schwarz contribution coming from the chiral tensor field on deformation classes of 7-manifolds $X$ associated to generators of the bordism group $\Omega^{\text{Spin}}_7 (B\mathbb{Z}_k)$. This is
\begin{equation}\label{eq:anomalycanc}
    \mathcal{A}^F = \kappa \, \widetilde{\mathcal{Q}} (\check{C}) \,,
\end{equation}
where $\kappa$ encodes the chirality, $\check{C}$ denotes a background field composed of the gauge field data, and $\widetilde{\mathcal{Q}}$ is the non-perturbative part of the quadratic refinement. One can read this equation as the requirement that the non-perturbative fermion anomaly $\mathcal{A}^F$ itself is a quadratic refinement of the gauge data and consequently has to satisfy its universal properties under change of background fields and Spin structure. Imposing these properties lead to strong restrictions for the allowed fermion spectra and the anomaly inflow onto strings, independent of the choice of the quadratic refinement. Once a particular choice is made, these constraints are further refined. We summarize our results in Table \ref{tab:results}. These constraints assume that the two contributions in \eqref{eq:anomalycanc} do not vanish on their own. Another valid solution is of course that the fermion anomaly vanishes on its own and the tensor field does not couple to the background.

\begin{table}
\resizebox{\textwidth}{!}{
\begin{tabular}{c || c | c |}
Group & Independent of quadratic refinement & Specific quadratic refinement \\ \hline \hline
$\mathbb{Z}_2$ & $n_1 = 4 \, \text{ mod } \, 8$ & $n_1 = 4 \kappa (1+2r) \, \text{ mod } \, 16$ \\ \hline
$\mathbb{Z}_3$ & $n_1 + n_2 = 0 \, \text{ mod } \, 3$ & $n_1 + n_2 = 3 \kappa r \, \text{ mod } 9$ \\ \hline 
$\mathbb{Z}_4$ & $n_1 + n_3 = 4j \, \text{ mod }\, 8$ & $5(n_1+n_3)+8n_2 = \begin{cases}
        \kappa(4 + 8r) \, \text{ mod }\, 32   &\text{ for $j$ odd}\\
        8 \kappa r \, \text{ mod }\, 32 &\text{ for $j$ even}
\end{cases}$ \\ 
 & $n_2 = 0 \, \text{ mod }\, 2$ & \\ \hline
$\mathbb{Z}_5$ & $n_1+n_2+n_3+n_4 = 3 \kappa j^2 \, \text{ mod } \, 5$ & $n_1 + n_4 + 2(n_2 + n_3) = \kappa r \text{ mod } \, 5$ \\
\hline
$\mathbb{Z}_6$ & $n_1 + n_3 + n_5 = 4 \, \text{ mod }\, 8$ &  $35(n_1+n_5)+80(n_2+n_4)+99n_3 = $\\ 
&$n_2 + n_3 + n_4 +n_5 = 0 \, \text{ mod }\, 3$ &  $=\begin{cases}
        \kappa(12 + 24r) \, \text{ mod }\, 144   &\text{ for $j$ odd}\\
        24 \kappa r \, \text{ mod }\, 144 &\text{ for $j$ even}
    \end{cases}$ \\
\hline
$\mathbb{Z}_7$ & $4 (n_1 + n_6) +  n_2 + n_5 +$ & $2(n_1 + n_6) + 5(n_2 + n_5) = \kappa r \text{ mod } \, 7$ \\ 
& $+ 2 (n_3 + n_4) =  \kappa j^2 \enspace \text{ mod } \, 7 $ & \\ \hline
\end{tabular}}
\caption{Summary of our results: $n_q$ counts the effective number of chiral fermions with charge $q$ mod $k$, $\kappa$ refers to the chirality of the 2-form field ($+1$ for self-dual), and $j \in \{ 1 \,, 2 \,, \dots \,, k-1\}$ is a parameter that links six-dimensional anomaly cancellation to the anomaly inflow onto the strings coupling to the 2-form fields. The parameter $r \in \{ 0 \,, 1 \,, \dots \,, k-1 \}$ parametrizes the choice of quadratic refinement. Here, here we assume that the fermion anomaly does not vanish on its own.}
\label{tab:results}
\end{table}

Since anomalies only depend on topological data of the theory, they are very robust against any continuous deformations of the theory. This includes supersymmetry breaking and our results are equally valid in non-supersymmetric theories in six dimensions. That being said, the setup of chiral tensors canceling anomalies of other sectors naturally appears in $\mathcal{N} = (1,0)$ supergravity theories with eight real supercharges, which contains chiral 2-forms in the gravity as well as the tensor multiplets, \cite{Green:1984bx, Sagnotti:1992qw}. Many of these supergravity theories, with potential exceptions \cite{Taylor:2018khc}, appear via compactification of F-theory on genus-one-fibered Calabi-Yau 3-folds \cite{Taylor:2011wt}, for which perturbative anomalies are encoded in the geometry of the compactification space \cite{Sadov:1996zm, Park:2011ji} and discrete gauge theories in multi-sections \cite{Morrison:2014era, Anderson:2014yva, Garcia-Etxebarria:2014qua, Mayrhofer:2014haa}. A version of this seems to hold also for the discrete anomalies \cite{Dierigl:2022zll} and our general bottom-up constraints are in agreement with the top-down construction of \cite{Dierigl:2022zll}, see also \cite{Oehlmann:2019ohh, Knapp:2021vkm, Schimannek:2021pau}, more of which will appear in \cite{DOT2025}.

The rest of the manuscript is organized as follows: In Section \ref{sec:anomalies} we review the anomalies of six-dimensional field theories, with a focus on non-perturbative anomalies in the presence of discrete gauge theories in Section \ref{sec:discanom}. In Section \ref{sec:const} we explain our strategy to obtain consistency constraints on fermion spectra from general properties of quadratic refinements. These are implemented for discrete gauge theories with gauge group $\mathbb{Z}_k$, with $k \in \{ 2, 3, 4, 5, 6, 7 \}$. Some more gauge groups are covered in Appendix \ref{app:higherk}. In Section \ref{sec:spin}, we incorporate further consistency constraints, which appear for even $k$ from a change of Spin structure when using the natural quadratic refinements built as in \cite{Hopkins:2002rd}. We conclude and point out interesting future directions in Section \ref{sec:concl}. Appendix \ref{app:bordprime} contains a proof for the bordism groups $\Omega^{\text{Spin}}_7 (B\mathbb{Z}_p)$, for $p$ an odd prime; Appendix \ref{app:spinQR} contains a proof of the change of the anomalies under a change of Spin structure. Some more technical details concerning quadratic refinements and fermion anomalies are summarized in Appendix \ref{app:redQR} and \ref{app:eta}.

\section{Anomalies for chiral fields}
\label{sec:anomalies}

In this section we will discuss the anomalies for chiral fields in a six-dimensional theory. Since anomaly constraints are topological, these considerations are independent of supersymmetry. Nevertheless, with an eye towards the study of $\mathcal{N}=(1,0)$ supergravity theories in six dimensions we include the chiral fields associated to spin-$\tfrac{1}{2}$ fermions, spin-$\tfrac{3}{2}$ fermions, as well as chiral tensor fields, all appearing naturally in 6d supergravity multiplets, see, e.g., \cite{Taylor:2011wt}.

Since we are particularly interested in anomalies involving discrete gauge symmetries we need to go beyond a perturbative discussion. The reason is that perturbative anomalies are associated with small gauge transformations which can be continuously connected to the identity. For discrete gauge theories, these do not exist and one instead needs to allow for gauge transformations that are not connected to the identity. Hence, the gauge and mixed gauge-gravitational anomalies are purely non-perturbative whereas the pure gravitational anomalies are perturbative and can be determined using infinitesimal diffeomorphisms.

This non-perturbative nature of the anomalies requires the use of the techniques developed in \cite{Dai:1994kq, Freed:2014iua, Witten:2015aba, Yonekura:2016wuc, Witten:2019bou}, see also \cite{Garcia-Etxebarria:2018ajm, Hsieh:2018ifc, Wan:2018bns}. In particular, one introduces an invertible theory, i.e., a theory with one-dimensional Hilbert space, in one higher dimension (here: 7), the anomaly theory $\mathcal{A}$ \cite{Freed:2012bs, Freed:2014iua}. It only contains information about the complex phase of the partition function of the theory of interest. To be more precise, evaluating the anomaly theory on a manifold $X$ with boundary $\partial X = M$ one has:
\begin{equation}
    \frac{Z [M]}{|Z[M]|} = e^{2 \pi i \, \mathcal{A}[X]} \,.
\end{equation}
Since anomalies are captured precisely by a difference in phase of the partition function evaluated on gauge equivalent backgrounds, the anomaly theory contains all information about anomalies. In cases where perturbative anomalies are absent, the anomaly theory becomes topological and depends only on the deformation class of the 7-manifold $X$. The non-perturbative anomalies are absent if
\begin{equation}
    e^{2 \pi i \, \mathcal{A}[X]} = 1 \,, \quad \text{ for all } \enspace  X \in \Omega^{\xi}_7 (BG) \,,
\end{equation}
where $\Omega^{\xi}_7 (BG)$ denotes the bordism group of deformation classes of closed 7-manifolds, including data on their tangent bundle, encoded in $\xi$, and host a gauge field for the gauge group $G$, with $BG$ denoting the classifying space.

\subsection{Fermion anomalies}
\label{subsec:fermanom}

The anomaly theory for fermionic fields are captured by so called $\eta$-invariants, see \cite{Atiyah:1975jf, Atiyah:1976jg, Atiyah:1976qjr, Dai:1994kq, Witten:2015aba, Yonekura:2016wuc}, which can be expressed as a regularized sum of the signs of the eigenvalues $\lambda_{\mathcal{D}}$ of a Dirac-type operator $\mathcal{D}$
\begin{equation}
    \eta^{\mathcal{D}} [X] = \tfrac{1}{2} \Big( \sum_{\text{reg}} \text{sign}(\lambda_{\mathcal{D}}) \Big) \,.
\end{equation}
Here, we will mainly need the Dirac operator of charged fermions, whose $\eta$-invariant we denote by $\eta^{\text{D}}$, and its analog for neutral spin-$\tfrac{3}{2}$ Rarita-Schwinger fields, denoted by $\eta^{\text{RS}}$. If one evaluates the $\eta$-invariant on a manifold $X$ which is itself a boundary $X = \partial Z$, i.e., $X$ is trivial in $\Omega^{\xi}_7(BG)$, the Atiyah-Patodi-Singer index theorem \cite{Atiyah:1975jf, Atiyah:1976jg, Atiyah:1976qjr} shows
\begin{equation}
    \text{Index}^{\mathcal{D}} = \int_{Z} I^{\mathcal{D}} + \eta^{\mathcal{D}}[X] \in \mathbb{Z}  \,.
    \label{eq:APS}
\end{equation}
Here, $\mathcal{D}$ is extended to $Z$ with APS boundary conditions. The gravitational part of the index densities $I^{\mathcal{D}}_8$ take the form
\begin{equation}
    \begin{split}
        I^{\text{D}}_8 &= - \tfrac{1}{5760} (7 p_1^2 - 4p_2) \,, \\
        I^{\text{RS}}_8 &= \tfrac{1}{720} (37 p_1^2 - 124 p_2) \,,
    \end{split}
    \label{eq:gravanomdens}
\end{equation}
which correspond to the respective contributions to the anomaly polynomial. Since we only consider discrete gauge theories, which do not have a continuous field strength, the gauge part in form of the Chern character $\text{ch}(F)$ is absent. The continuous dependence on the metric is given in terms of Pontryagin classes $p_i$ whose Chern-Weil representatives are given by
\begin{equation}
    p_1 = - \tfrac{1}{2} \text{tr}(R^2) \,, \quad p_2 = \tfrac{1}{8} \text{tr} (R^2)^2 - \tfrac{1}{4} \text{tr}(R^4) \,,
    \label{eq:CWpn}
\end{equation}
where we absorbed a factor of $\tfrac{1}{2 \pi}$ in the definition of the curvature 2-form $R$.

Once the perturbative gravitational anomalies are canceled, we can work with reduced $\eta$-invariants for fermions of charge $q$ under the $\mathbb{Z}_k$ gauge group, for which the pure gravitational part is subtracted
\begin{equation}
\widetilde{\eta}^{\text{D}}_q = \eta^{\text{D}}_q - \eta^{\text{D}}_0 \,.
\label{eq:redeta}
\end{equation}
The fermion contribution to the anomaly theory for the non-perturbative gauge anomalies is therefore given by
\begin{equation}
\mathcal{A}^F = \sum_{q = 1}^{k-1} n_q \, \widetilde{\eta}^{\text{D}}_q \,,
\label{eq:fermanomtheo}
\end{equation}
where the $n_q$ are the multiplicities. Motivated by supergravity applications, we will consider charged fermions of fixed chiralities, as expected from the hypermultiplet sector.\footnote{For different chiralities $n_q$ has contributions with both signs.} We further do not include charged spin-$\tfrac{3}{2}$ fermions and the non-perturbative part of the fermion anomaly is determined by the Weyl fermion spectrum in six dimensions. As discussed above, the theory $\mathcal{A}^F$ defined in this way is topological and only depends on the deformation class of the underlying 7-manifold. In particular, one has
\begin{equation}
    e^{2 \pi i \, \mathcal{A}^F [\partial Z]} = 1 \,,
\end{equation}
for boundaries.

\subsection{Tensor anomalies}
\label{subsec:tenanom}

Similar to chiral fermions also chiral tensor fields contribute to the anomalies. In six dimensions the chirality is encoded in the (anti-)self-duality of the 2-form fields $B$, which locally takes the form $F_B = d B$,
\begin{equation}
\ast F_B = \kappa F_B \,,
\end{equation}
with $\kappa = 1$ for self-dual and $\kappa = -1$ for anti-self-dual fields. This theory can be expressed as the boundary of a seven-dimensional theory with 3-form field $A$, whose boundary mode gives rise to the (anti-)self-dual $B$, see, e.g., \cite{Witten:1996hc, Belov:2006jd, Hsieh:2020jpj}. The relevant part\footnote{For the evaluation of the anomaly contribution the kinetic term is not important.} of the boundary action is encoded in a Chern-Simons like term of the form $A \wedge dA$ potentially coupled to a background $C$. To be well-defined, this Chern-Simons term requires some explanation.

First of all, in order to treat topologically non-trivial configurations for the 3-form field $A$ a description in terms of a differential form is not sufficient. Instead, one has to encode the 3-form field in terms of differential cohomology, see, e.g., \cite{Hopkins:2002rd, Freed:2006yc, Hsieh:2020jpj}, or more precisely an element in $\check{H}^4 (X)$. The information in such an element in differential cohomology will be described below.

Second, the Chern-Simons term also appears at a half-integer level. This means that more refined structures of the underlying spacetime manifold enter the contributions of the chiral tensor fields. These can be phrased in terms of a quadratic refinement $\mathcal{Q}$ of the differential cohomology pairing
\begin{equation}
\check{H}^4 (X) \times \check{H}^4 (X) \rightarrow \text{U}(1) \,.
\end{equation}
A careful analysis, see \cite{Hsieh:2020jpj}, shows that the gravitational part of the anomaly theory of a chiral tensor field is given by
\begin{equation}
\mathcal{A}^B_{\text{grav}} = 28 \kappa \, \eta^{\text{D}}_0 \,.
\end{equation}
Additionally, one has the contribution induced by the coupling to the background $\check{C}$ now also lifted to a class in $\check{H}^4(X)$, given by
\begin{equation}
\mathcal{A}^B_{\text{GS}} = - \kappa \, \widetilde{\mathcal{Q}} (\check{C}) \,,
\end{equation}
where in our case $\widetilde{\mathcal{Q}}$ denotes a reduced quadratic refinement only sensitive to the non-perturbative anomaly contribution, in analogy to \eqref{eq:redeta}. Choosing the background field $\check{C}$ appropriately in terms of the characteristic classes of the gauge and tangent bundle, such that one has
\begin{equation}
\big( \mathcal{A}^F + \mathcal{A}^B_{\text{GS}} \big) [X] \in \mathbb{Z} \,, \quad \text{for all } X \in \Omega^{\text{Spin}}_7 (B \mathbb{Z}_k) \,,
\end{equation}
amounts to a non-perturbative version of the Green-Schwarz anomaly cancellation mechanism.\footnote{Note that in supergravity the quadratic Green-Schwarz mechanism needs to contain a perturbative gravitational piece in order to cancel part of the gravitino anomaly.}

Before we go to the investigation of discrete anomaly cancellation for $\mathbb{Z}_k$ gauge theories let us introduce the differential cohomology description and the properties of a quadratic refinement more explicitly.

\subsection{Differential cohomology and discrete gauge fields}
\label{subsec:difgauge}

In the following, we will focus on the details needed to understand the description of chiral $p$-form fields using differential cohomology and refer to \cite{Hopkins:2002rd, Freed:2006yc,Monnier:2016jlo,Hsieh:2020jpj} for more details.

Differential cohomology is a very convenient way to express the gauge invariant information contained in a U$(1)$ $(p-1)$-form field. It is described by a differential character, i.e., a representative of an element in $\check{H}^{p} (X)$, where $X$ refers to the spacetime manifold. Important for us is that the differential character does not only know about the field strength but also the gauge connection and its holonomies, thus containing more refined information. It is convenient to describe a differential character $\check{C}$ as a triple $(N_C,A_C,F_C)$, whose individual parts we are going to explain next.

The piece $F_C$ refers to the field strength, which is a closed $p$-form, which is gauge invariant. Whenever $F_C$ is zero everywhere, we call the associated differential character flat. $F_C$ is naturally a representative of an integer-quantized de Rham cohomology class $H^{p}(X;\mathbb{R})$. However, a differential character can have more refined topological data as an element in $H^{p}(X;\mathbb{Z})$, for which the characteristic class $N_C$ is a representative. Whenever $N_C$ vanishes the differential character is called topologically trivial.\footnote{Note that there are flat fields that are topologically non-trivial, as well as non-flat fields that are topologically trivial.} Essentially $N_C$ parameterizes the failure of the connection $A_C$ to be a differential form. The connection $A_C$, satisfies the equation
\begin{equation}
    \delta A_C = F_C - N_C \,,
\end{equation}
and should be understood as a map 
\begin{equation}
    A_C: \quad C_{p-1}(X) \rightarrow \mathbb{R} \,,
\end{equation}
from the $(p-1)$-chains of $X$ to the real numbers. This also defines the holonomy function
\begin{equation}
    \chi (M_{p-1}) = e^{2 \pi i \int_{M_{p-1}} A_C} \,,
    \label{eq:holfunc}
\end{equation}
defined for closed $(p-1)$-dimensional submanifolds $M_{p-1}$, which features in alternative description of differential cohomology.

The fact that we are dealing with a cohomology theory implies that there is some equivalence relation, which on the individual entries of the differential character acts as
\begin{equation}
    N_C \rightarrow N_C - \delta n \,, \quad A_C \rightarrow A_C + \delta a + n \,,
    \label{eq:gaugedifcohom}
\end{equation}
with $F_C$ being gauge invariant. Here, $n$ is an integer $(p-1)$ co-chain and $a$ a real $(p-2)$ co-chain, which morally can be understood as large and small gauge transformations of the connection $A_C$, respectively.
\\One can multiply differential characters
\begin{equation} \label{eq:diffcohoprod}
    *: \quad \check{H}^{p}(X) \times \check{H}^q (X) \rightarrow \check{H}^{p+q} (X) \,.
\end{equation}
The resulting field strength and characteristic class of $\check{C}*\check{B}$ are given by
\begin{equation}
    F_{C*B} = F_C \wedge F_B \,, \quad N_{C*B} = N_C \cup N_B \,,
\end{equation}
whereas the connection piece is a little more complicated and can be understood as a refined version of a Chern-Simons term $A_C \wedge d A_B$, see, e.g., \cite{Hsieh:2020jpj}. If either character is topologically trivial the Chern-Simons term is the correct answer.

Importantly, differential characters $\check{C} \in \check{H}^{p}(X)$ can be integrated over both $p$-dimensional manifolds and $(p-1)$-dimensional ones, see also \cite{GarciaEtxebarria:2024fuk}. The integral over a closed $(p-1)$-dimensional manifold can be understood as a map to U$(1)$ given precisely by the holonomy function \eqref{eq:holfunc}. Together with the product \eqref{eq:diffcohoprod} it defines a natural U$(1)$-valued pairing on closed $d$-dimensional manifolds
\begin{equation}
\begin{split}
    (\cdot\,,\cdot):& \quad \check{H}^p (X) \times \check{H}^{d-p+1} (X) \mapsto \text{U}(1) \,, \\
    & \quad (\check{C}, \check{B}) \mapsto \int_{X} A_{C * B} \,.
    \label{eq:difcohompairing}
\end{split}
\end{equation}
When both fields are flat, i.e., their field strength vanishes, this pairing coincides with the torsion pairing in singular cohomology. For odd $d$ this torsion pairing takes the form
\begin{equation}
    (\cdot\,,\cdot): \text{Tor} \big(H^{(d+1)/2} (X;\mathbb{Z}) \big) \times \text{Tor} \big(H^{(d+1)/2} (X;\mathbb{Z}) \big) \rightarrow \text{U}(1) \,,
\end{equation}
defined as follows: let $\alpha, \beta$ be a representatives of classes in $\text{Tor}\big( H^{(d+1)/2} (X;\mathbb{Z}) \big)$. Since $\alpha$ is torsion one has
\begin{equation}
    n \alpha = \delta \sigma \,, \quad \sigma \in C^{(d-1)/2} (X_d;\mathbb{Z}) \,,
\end{equation}
and the torsion pairing is defined as
\begin{equation}
    (\alpha,\beta) = \frac{1}{n} \int_{X} \sigma \cup \beta \,.
\end{equation}
A dual pairing can be defined on the level of homology for two torsion $(\tfrac{d-1}{2})$-cycles and is often referred to as the linking pairing, see, e.g., \cite{GarciaEtxebarria:2019caf}.

\subsubsection*{Discrete gauge fields and differential cohomology}

In the main part of this work we will discuss gauge theories with 0-form gauge group $\mathbb{Z}_k$. These do not have any non-trivial field strength and are therefore described by flat differential characters $\check{c}$ in $\check{H}^2 (X)$, which we will denote by $\check{H}^2_{\text{flat}} (X)$. The characteristic class of such a flat character is in general non-trivial, as is its connection. In particular, a $\mathbb{Z}_k$ gauge field is fully specified by maps
\begin{equation}
    H_1 (X; \mathbb{Z}) \rightarrow \mathbb{Z}_k \,,
\end{equation}
or in other words, elements in $H^1 (X;\mathbb{Z}_k)$. Using the natural embedding $\mathbb{Z}_k \hookrightarrow \text{U}(1)$ this can be interpreted as an element $H^1\big(X;\text{U}(1)\big)$. Indeed, the space of flat characters $\check{H}_{\text{flat}}^p (X)$ is naturally isomorphic to $H^{p-1}\big(X;\text{U}(1)\big)$, which for $p =2$ reduces to the case of the discrete gauge fields.\footnote{For a flat character $\delta A = -N$, so if we take the reduction of $A$ to U$(1)\simeq\R/\Z$ coefficients, we get a U$(1)$ valued co-cycle with $\delta A_{\text{U}(1)} = 0$. Since it is defined up to gauge equivalence, it gives a class $[A]_{\text{U}(1)} \in H^{p-1}\big(X;\text{U}(1)\big)$. In the other direction, take a closed co-chain $A_{\text{U}(1)}$ and arbitrarily lift it to a real co-chain $A_\R$. Then define the characteristic class as $N = -\delta A_\R$. This is nothing but the Bockstein map $\beta$, so we obtained a flat character of the form $(\beta(A),A,0)$. If one is careful about the ambiguities in the definition of a co-chain representing $\beta(A),$ they are exactly those of the gauge ambiguity in \eqref{eq:gaugedifcohom}, see also \cite{Hsieh:2020jpj}.\label{footnote:flatchar}} The characteristic class can be derived by the Bockstein homomorphism associated to the short exact sequence
\begin{equation}
    0 \rightarrow \mathbb{Z} \rightarrow \mathbb{R} \rightarrow \text{U}(1) \rightarrow 0 \,,
\end{equation}
which maps
\begin{equation}
    \beta: \quad H^1 \big(X;\text{U}(1)\big) \rightarrow H^2(X; \mathbb{Z}) \,.
\end{equation}
The image of the connection $A_c$, interpreted as an element of $H^1(X;\text{U}(1))$, precisely gives the characteristic class of the flat differential character $\big(\beta(A_c),A_c,0\big)$. Higher cohomology classes can be constructed using the product defined above. For anomaly cancellation, we will use an element in $\check{H}^4(X)$ obtained as $\check{c}*\check{c}$, which is again a flat character since the field strength vanishes. Its connection piece is given by
\begin{equation}
    A_{c * c} = A_c \cup N_c = A_c \cup \beta(A_c) \,,
\end{equation}
where we also used the natural multiplication $\mathbb{R} \times \mathbb{Z} \rightarrow \mathbb{R}$ and note that due to the large gauge transformations its integral takes values in U$(1)$ as expected. Thus, one has
\begin{equation}
\check{c} \ast \check{c} = \big(\beta(A_c)\cup \beta(A_c) \,, A_c\cup \beta(A_c) \,,0 \big) \,,
\end{equation}
as an element in $\check{H}^4_{\text{flat}}(X)$ constructed out of the gauge background.

\subsubsection*{Gravitational background and differential cohomology}

In the gravitational sector things are more general, since the fields are not necessarily flat. The main object we need to consider is a differential refinement of a class $\lambda \in H^4(X;\mathbb{Z})$, which is defined by
\begin{equation}
    2 \lambda = p_1 \,,
\end{equation}
and is well-defined on Spin manifolds.\footnote{This can be seen from pulling back the  generating element $H^4 (B\text{Spin};\mathbb{Z})$ to $X$.} We will mainly denote it as $\tfrac{1}{2} p_1$. The characteristic class and field strength of this differential character are given by
\begin{equation}
    N_g = \tfrac{1}{2} p_1 \,, \quad F_g = - \tfrac{1}{2} \text{tr} (R^2)
\end{equation}
where we use subscript $_g$ to refer to the gravitational origin. The connection satisfies
\begin{equation}
    \delta A_g = F_g - N_g \,,
\end{equation}
and completes the differential character $\check{\lambda} = \tfrac{1}{2}\check{p}_1 = (N_g, A_g, F_g) \in \check{H}^4 (X)$.

\subsection{Quadratic refinements}
\label{subsec:QR}

We mentioned in Section \ref{subsec:tenanom}, that the definition of the chiral 2-form fields require a quadratic refinement $\mathcal{Q}$. For two Abelian groups $G$ and $H$ with a bilinear pairing
\begin{equation}
    (.\,,.): \quad G \times G \rightarrow H \,,
\end{equation}
a quadratic refinement is a function
\begin{equation}
    \mathcal{Q}: \quad G \rightarrow H \,, 
\end{equation}
such that for all $x,y \in G$ one has
\begin{equation}
    (x,y) = \mathcal{Q}(x+y) - \mathcal{Q}(x) - \mathcal{Q}(y)+\mathcal{Q}(0) \,.
    \label{eq:QRprop}
\end{equation}
For $G = H = \mathbb{R}$ and the product given by usual multiplication a valid choice of a quadratic refinement is given by $\mathcal{Q}(x) = \tfrac{x^2}{2}$, so \eqref{eq:QRprop} essentially implements a division by two. More interesting cases of quadratic refinements arise in situations where such a division by two is not well-defined. 

To illustrate this, let us consider $G = H = \mathbb{Z}$ with the bilinear product given again by multiplication. Within the integers one cannot generally divide by two, which means that a quadratic refinement must take a more general form. The simplest function satisfying \eqref{eq:QRprop} is given by $\mathcal{Q}(x) = \tfrac{1}{2} x (x+1)$. As simple as this may seem, the quadratic refinements in the cases of interest to us are of the same type, where $(\cdot,\cdot)$ arises as pairing of (torsional) integer cohomology classes.

On a closed oriented 8-manifold $Z$ there is a natural pairing in singular cohomology, given by
\begin{equation}
    (x,y) = \int x \cup y \,, \quad x \,, y \in H^4(Z;\mathbb{Z}) \,.
\end{equation}
A quadratic refinement for this pairing is provided by the fourth Wu class $\nu_4 \in H^4(Z;\mathbb{Z}_2)$, for which we define an integer lift $\nu_4^{\mathbb{Z}} \in H^4(Z;\mathbb{Z})$, which reduces to $\nu_4$ mod $2$. This class has the property that for every $x \in H^4(Z;\mathbb{Z})$\footnote{The existence of such a class can be shown fairly easily: With $\mathbb{Z}_2$ coefficients squaring $x \mapsto x^2 = x \cup x$ is a linear function, since $(x+y)^2 = x^2 + y^2$. By Poincar\'e duality it can be represented by the cup product with a class $\nu_4 \in H^4(Z;\mathbb{Z}_2)$. Any integer lift of $\nu_4$ satisfies \eqref{eq:characelem}. The uniqueness of the mod 2 class is non-trivial but can be shown to be true, see, e.g., \cite{MR1867354}.}
\begin{equation}
    \int x \cup \nu_4^{\mathbb{Z}} = \int x \cup x \enspace \text{mod }2 \,.
\label{eq:characelem}
\end{equation}
Thus, a quadratic refinement of the pairing is given by
\begin{equation}
    \mathcal{Q}(x) = \tfrac{1}{2} \int x \cup (x + \nu_4^{\mathbb{Z}})
\end{equation}
Here, $\nu_4^{\mathbb{Z}}$ plays the same role as $1$ above, being a characteristic element of the pairing, i.e., satisfying \eqref{eq:characelem}, see also the discussion in \cite{Monnier:2018nfs}.

For Spin manifolds, with first and second Stiefel-Whitney class trivial, one has that $\nu_4 = w_4$ and an integer lift is provided by
\begin{equation}\label{eq:wu4}
    \nu_4^{\mathbb{Z}} = - \tfrac{1}{2} p_1 = -\lambda \,.
\end{equation}
Adding an even class in $H^4(Z;\mathbb{Z})$ will not change \eqref{eq:characelem} so we can construct quadratic refinements using any class of the form
\begin{equation}
    \tilde{\nu}_4^{\mathbb{Z}} = - \frac{2l + 1}{2} p_1 \,.
    \label{eq:genWu}
\end{equation}
These classes can be lifted to elements in differential cohomology, similar to $\tfrac{1}{2} p_1$ discussed above and we will denote the corresponding differential characters as $\check{\nu}$.

Our main concern will not be a quadratic refinement of the pairing on an 8-manifold, but rather a quadratic refinement for the pairing of elements in $\check{H}^4(X)$ on a Spin 7-manifold, defined in \eqref{eq:difcohompairing}. In \cite{Hsieh:2020jpj} an explicit construction for a specific quadratic refinement was provided, which we will briefly recall in the following. In particular this choice of the quadratic refinement relies on the extension to a Spin 8-manifold $Z$, such that $\partial Z = X$, over which the background field $\check{C}$ interpreted as a U$(1)$ 3-form gauge field extends. This is always possible since $\Omega^{\text{Spin}}_7 \big( K(\mathbb{Z};4)\big) = 0$, \cite{Stong}. The quadratic refinement is given by
\begin{equation}
    \mathcal{Q}(\check{C}) = \int_Z \Big( \frac{1}{2} F_C \wedge F_C - \frac{1}{4} p_1 \wedge F_C + 28 \, I^{\text{D}}_8\Big) \enspace \text{ mod } \, 1 \,,
    \label{eq:QRTachikawa}
\end{equation}
where $I^{\text{D}}_8$ follows from the $\hat{A}$ genus as given in \eqref{eq:gravanomdens}. It is easy to check that this satisfies \eqref{eq:QRprop}. Moreover it is independent of the extension to $Z$, which can be demonstrated as in \cite{Hsieh:2020jpj} via index theory, or via the property \eqref{eq:characelem} of the Wu class.

In this work the quadratic refinement depends on a choice of background in $\check{H}^4(X)$, which can depend on gauge an gravitational data. Specifically we want to isolate the part sensitive to the discrete gauge theory and its non-perturbative anomalies and define
\begin{equation}
    \widetilde{\mathcal{Q}} (\check{C}) = \mathcal{Q}(\check{C}) - \mathcal{Q}_0 \,,
    \label{eq:redQ}
\end{equation}
where $\mathcal{Q}_0$ is given by $\mathcal{Q}(\check{C})$ with the gauge bundle switched off. Note, that in general this differs from $\mathcal{Q}(0)$, used in \cite{Hsieh:2020jpj}, since $\check{C}$ can contain gravitational contributions. These are necessary if there is a remnant perturbative gravitational anomaly that needs to be canceled, e.g., in supergravity where gravitini are present. We demonstrate in Appendix \ref{app:redQR} that $\widetilde{\mathcal{Q}}$ still satisfies \eqref{eq:QRprop} of a quadratic refinement. Since this only depends on the gauge background it can be also written in terms of $\check{C} \sim \check{c}*\check{c}$ as an element in $\check{H}^4(X)$. Note that this construction ensures that $\widetilde{\mathcal{Q}}$ is bordism invariant and therefore only depends on the deformation class of the 7-manifold $X$.

\vspace{0.3cm}

With all these definitions and general properties in place we can now analyze the anomalies of six-dimensional gravitational theories with gauge group $\mathbb{Z}_k$.

\section{Discrete gauge anomalies}
\label{sec:discanom}

In the present work, we will restrict to spacetime manifolds that admit a Spin structure\footnote{It would interesting to extend our discussion to situations where the gravitational and gauge backgrounds mix, such as for example Spin-$\mathbb{Z}_{2k}$ structures, see \cite{Garcia-Etxebarria:2018ajm, Wan:2018bns, Debray:2021vob, Debray:2023yrs}.} and have a single discrete gauge factor $\mathbb{Z}_k$, thus the bordism groups of interest to us are given by
\begin{equation}
    \Omega^{\text{Spin}}_7 (B \mathbb{Z}_k) \,.
\end{equation}
These classify the potential discrete anomalies of the theory, which include both pure gauge as well as mixed gauge-gravitational contributions.

\subsection{Bordism groups and generators}
\label{subsec:bordgen}

Some of the bordism groups $\Omega^{\text{Spin}}_7 (B\mathbb{Z}_k)$, can be found in the literature, see for example \cite{Garcia-Etxebarria:2018ajm, Hsieh:2018ifc, Wan:2018bns} and references therein. In the following we will restrict to the cases $2 \leq k \leq 7$, each of which give something new and interesting, see Appendix \ref{app:bordprime} for a derivation of the bordism group for $k = p$ being any odd prime.

In the range of interest the bordism groups and their generators are given by
\begin{equation}
    \begin{array}{ c || c | c |}
    k & \Omega^{\text{Spin}}_7 (B\mathbb{Z}_k) & \text{generators} \\ \hline \hline
    2 & \mathbb{Z}_{16} & L^7_2 \\ 
    3 & \mathbb{Z}_9 & L^7_3 \\ 
    4 & \mathbb{Z}_{32} \oplus \mathbb{Z}_2 & L^{7}_4 \,, \widetilde{L}^7_4 \\ 
    5 & \mathbb{Z}_5 \oplus \mathbb{Z}_5 & L^7_5, L^3_5 \times \text{K3} \\ 
    6 & \mathbb{Z}_{16} \oplus \mathbb{Z}_{9} & L^7_6 \\
    7 & \mathbb{Z}_7 \oplus \mathbb{Z}_7 & L^7_7 \,, L^3_7 \times \text{K3} \\ \hline
    \end{array}
    \label{eq:bordgen}
\end{equation}
The generators $L^n_k$, with $n$ odd, are lens spaces defined in the following way: consider $\tfrac{n+1}{2}$-dimensional complex space $\mathbb{C}^{(n+1)/2}$, with coordinates $z_i$, acted upon by the $\mathbb{Z}_k$ action
\begin{equation}
    \mathbb{Z}_k: \quad z_i \mapsto e^{2 \pi i / k} z_i \,.
\end{equation}
This acts freely on $S^n$, at asymptotic infinity (or fixed radius), whose quotient gives the lens space $L^n_k = S^n/\mathbb{Z}_k$. For $n \in \{ 3, 7\}$ all the lens spaces are Spin. For even $k$ they admit two Spin structures, which we distinguish by $L^n_k$ and $\widetilde{L}^n_k$.

The homology groups of $L^n_k$ with integer coefficients are given by
\begin{equation}
     H_i (L^{n}_k ; \mathbb{Z}) =
        \begin{cases}
            \mathbb{Z} \,, &\text{ for } i \in \{0 \,, n\} \,, \\
            \mathbb{Z}_k \,, &\text{ for } i \in \{ 1 \,, 3 \,, \dots \,, n-2 \} \,, \\
            0 \,, &\text{ otherwise} \,.
        \end{cases}
\end{equation}
Using the universal coefficient theorem, this translates to cohomology groups with $G$ coefficients
\begin{equation} \label{eq:lenscoho}
    H^i (L^{n}_k; G) =
        \begin{cases}
            G \,, & \text{ for } i \in \{ 0, n\} \,, \\ \text{ker}(G\xrightarrow[]{k}G) \,, \ & \text{ for } i \in \{ 1, 3,... , n-2\} \,, \\ G/kG \,, &\text{ for } i \in \{ 2, 4,... , n-1\} \,, \\0 \,,  &\text{ otherwise} \,.
        \end{cases}
\end{equation}
In particular, if $G = \mathbb{Z}_k$ one has that the cohomology groups are $H^i (L^n_k; \mathbb{Z}_k) = \mathbb{Z}_k$ for $i \in \{ 0 \,, 1 \,, \dots \,, n \}$ and vanish otherwise. For $G=\Z$ one has $\Z_k$ in even degrees, and zero otherwise, while for $G=\text{U}(1)$ one has $\Z_k$ in odd degrees, and zero otherwise.

From this construction one can also calculate the first Pontryagin class of the seven-dimensional lens spaces, which as an element of $H^4(L^7_k;\mathbb{Z})$ is torsion. Using the definition of the lens space as quotient of a sphere embedded in complex space, we find that the first Pontryagin class is given by
\begin{equation}
    p_1 (L^7_k) = 4 x^2 \,,
\end{equation}
with $x \in H^2(L^7_k ; \mathbb{Z})$ the natural generator.\footnote{The class $x$ can be understood as first Chern class of $z_i$ as interpreted as complex line bundle over $L^n_k$.}

The torsion pairing on the seven-dimensional lens spaces is given by
\begin{equation}
    (x^2,x^2) = - \frac{1}{k} \,,
\end{equation}
where $x^2 \in H^4(L^7_k;\mathbb{Z})$ is a natural generator constructed from $x \in H^2(L^7_k;\mathbb{Z})$. This can derived from a different construction of the lens space as a circle bundle over $\mathbb{CP}^{(n-1)/2}$ with the fiber given by the circle bundle in $\mathcal{O}(-k)$ at fixed radius, see \cite{Hsieh:2020jpj}, which leads to the same manifold $S^7/\mathbb{Z}_k$ above. This formulation has the advantage that one immediately obtains a bounding 8-manifold $Z$ by an extension to the disc bundle over $\mathbb{CP}^{(n-1)/2}$.

As mentioned above the discrete $\mathbb{Z}_k$ bundles are fully specified by an element in $H^1(X;\mathbb{Z}_k)$, which for the lens spaces is given by
\begin{equation}
    H^1(L^n_k;\mathbb{Z}_k) = \mathbb{Z}_k \,.
\end{equation}
Using the Bockstein $\beta_k$ associated to the short exact sequence
\begin{equation}
    0 \rightarrow \mathbb{Z} \rightarrow \mathbb{Z} \rightarrow \mathbb{Z}_k \rightarrow 0 \,,
\end{equation}
this uniquely relates this to a class in $H^2(L^7_k; \mathbb{Z})$. The gauge bundle is therefore specified by an integer $m$ modulo $k$, and its topological class is given by $m \, x \in H^2(L^7_k; \mathbb{Z})$. We denote the lens space with this bundle by $L^7_k(m)$ with $m =1$ also referred to as $L^7_k$. With the definitions in Section \ref{subsec:difgauge}, this gives rise to a flat differential character in $\check{H}^2_{\text{flat}}(L^7_k)$, which we denote as $\check{c}$, fully specified by
\begin{equation}
    [N_c] = m x \,.
\end{equation}
The differential lift of the gauge background, $\check{C} \in \check{H}^4_{\text{flat}}(L_k^n)$, entering the quadratic refinement then takes the form
\begin{equation}
    \check{C} = (m^2 x^2 \,, m^2 A_c \cup x \,, 0) \,,
\end{equation}
and is quadratic in $m$.

Also the inclusion of the gauge bundle has a description in terms of a circle bundle over $\mathbb{CP}^{(n-1)/2}$ after the $\mathbb{Z}_k$ bundle is embedded into a U$(1)$ bundle, which can be extended to $Z$ since $\Omega^{\text{Spin}}_7 \big( B\text{U}(1) \big) = 0$. As described in \cite{Hsieh:2020jpj} this can be understood as the pullback of $\mathcal{O}(-1)$ to the total space of $\mathcal{O}(-k)$, which corresponds to $Z$. Note that while the U$(1)$ bundle restricted to the boundary $\partial Z = L^n_k$ is flat, the same is not true in the interior.\footnote{This is precisely the reason why the U$(1)$ bundle can be extended to $Z$, while the necessarily flat $\mathbb{Z}_k$ bundle cannot.} The boundary is the same as the lens space $L^7_k(1)$ as constructed above. For the other gauge bundles one simply pulls back $\mathcal{O}(-m)\simeq \mathcal{O}(-1)^{\otimes m}$ to the total space of $\mathcal{O}(-k)$.

\subsection{Discrete fermion anomalies}
\label{subsec:discfermlens}

The discrete fermion anomalies are captured by the reduced $\eta$-invariants, defined in \eqref{eq:redeta}. For the lens spaces with discrete gauge bundle above these $\eta$-invariants can be determined using the equivariant index theorem, see \cite{f715b616-033c-3b18-9eb8-27d3c8f4674d, Hsieh:2020jpj}. For spin-$\tfrac{1}{2}$ fermions they read, see also Appendix \ref{app:eta},
\begin{equation}
    \widetilde{\eta}^{\text{D}}_q [L^n_k(m)] = (\eta^{\text{D}}_q - \eta^{\text{D}}_0) [L^7_k (m)]= - \frac{1}{k (2i)^{(n+1)/2}} \sum_{j = 1}^{k-1} \bigg( \frac{e^{- 2 \pi i j m q / k} - 1}{\big( \text{sin}(\pi j/k)\big)^{(n+1)/2}} \bigg) \,.
\end{equation}
From this formula it is easy to see that the reduced $\eta$-invariant vanishes whenever $mq$ is divisible by $k$.

For $k$ even we have the possibility to change the Spin structure on $L^7_k(m)$, which amounts to an additional minus sign for fermions traversing the torsion 1-cycle. This interpretation allows one to determine the $\eta$-invariants on these spaces by
\begin{equation}
    \widetilde{\eta}^{\text{D}}_q [\widetilde{L}^n_k(1)] =  (\eta^{\text{D}}_q - \eta^{\text{D}}_0) [\widetilde{L}^n_k(1)] = (\eta^{\text{D}}_{q+ \frac{k}{2}} - \eta^{\text{D}}_{\frac{k}{2}}) [L^n_k (1)] \,,
    \label{eq:etaspin}
\end{equation}
where the minus sign is implemented via a shift of the charges, see also the discussion in \cite{Debray:2021vob}.

Finally, we want to determine the $\eta$-invariants on product spaces as appearing in the list \eqref{eq:bordgen}. This is done using a decomposition in terms of index and $\eta$-invariants for the individual factors, see \cite{Atiyah:1976qjr, Gilkey1989TheGO}. For the Weyl fermion in the theory it reads
\begin{equation}
    \widetilde{\eta}^{\text{D}}_q [L^3_k \times \text{K3}] = \text{Index}^{\text{D}}_q [L^3_k] \, \widetilde{\eta}^{\text{D}}_q [\text{K3}] + \text{Index}^{\text{D}}_q [\text{K3}] \, \widetilde{\eta}^{\text{D}}_q [L^3_k] = 2 \, \widetilde{\eta}^{\text{D}}_q [L^3_k] \,,
    \label{eq:etaprod}
\end{equation}
where we used that $\text{Index}^{\text{D}}_q [\text{K3}] =2$ for any $q$, since there is no non-trivial discrete gauge bundle on the simply-connected K3.

With these explicit formulae we can calculate the discrete fermionic anomalies \eqref{eq:fermanomtheo} for all generators in \eqref{eq:bordgen} as well as different gauge bundles and Spin structures.

\subsection{Green-Schwarz contribution}
\label{subsec:discGSlens}

To specify the non-perturbative Green-Schwarz contribution
\begin{equation}
    \mathcal{A}^B_{\text{GS}} [X] = - \kappa \, \widetilde{\mathcal{Q}} (\check{C})[X] \,,
\end{equation}
we need to make some choices that specify the quadratic refinement. As in \cite{Monnier:2018nfs, Hsieh:2020jpj}, and recalled in Section \ref{subsec:QR}, one can use an extension to an 8-manifold $Z$, after embedding the discrete bundle into a U$(1)$ bundle. This 8-manifold, together with the extended U$(1)$ bundle, is given by the pull-back of $\mathcal{O}(-m)$ to the total space of $\mathcal{O}(-k)$. A generalized form of the quadratic refinement should then have an expression similar to \eqref{eq:QRTachikawa}.

Indeed, the quadratic refinements associated to the cancelable part of the non-perturbative fermion anomaly can be phrased in  precisely such a form and can be written as
\begin{equation}
    \widetilde{\mathcal{Q}} (j \check{C}) = \int_Z \Big( \frac{j^2}{2} F_C \wedge F_C - \frac{(2l+1)j}{4} p_1\wedge F_C \Big) \enspace \text{ mod } \, 1 \,,
    \label{eq:QRexpform}
\end{equation}
where the gauge background reduces to
\begin{equation}
    \check{C} = \check{c} \ast \check{c} = m^2 x^2 \,,
\end{equation}
on the boundary, with its non-flat U$(1)$ extension in the interior. Since we always express $\check{C}$ in terms of $x^2$, we will also use the notation $\widetilde{\mathcal{Q}}(N)$ with $N$ an integer mod $k$. This form might not be the most general, but is able to recreate the desired values on $L^7_k(m)$ and is bordism invariant.\footnote{To see that they do not depend on the extension, we can glue two different extensions along their common boundary to a closed 8-manifold and use the fact that $\tfrac{2l+1}{2} p_1$ is a characteristic element for the cup product, i.e., the integral \eqref{eq:QRexpform} evaluates to an integer.}

The parameter $j \in \{0\,,1\,,\dots\,,k-1\}$ is related to a coupling constant and is part of the defining data of the theory. This can be understood from the Green-Schwarz term in six dimensions, which morally takes the form
\begin{equation}
\mathcal{L}_{\text{GS}} \sim B \cup j N_C \,.
\end{equation}
It therefore specifies the gauge background seen by the 2-form fields. Due to the self-duality conditions it also fixes the anomaly inflow onto the strings coupling to $B$. It therefore relates the six-dimensional anomalies to worldsheet anomalies of a string. In the case $j=0$, the 2-forms do not couple to the background at all, and therefore the fermion anomaly should identically vanish on its own.

The second parameter $l \in \{0\,,1\,,\dots\,,k-1\}$ is also defined modulo $k$ and parametrizes the choice of lift of the Wu class to integer cohomology \eqref{eq:genWu} and consequently affects its differential lift. This leads to $k$ different choices for the quadratic refinement which can be specified by their value on the generator of $H^4(L^7_k;\mathbb{Z})$, i.e.,  $\widetilde{\mathcal{Q}}(1)$.

For this particular realization of the bounding manifold $Z$ we can also derive explicit formulae for the the quadratic refinement for the set of generators we are interested in, which follows the same lines as Appendix C of \cite{Hsieh:2020jpj}. In particular one finds that the first Pontryagin class is given by
\begin{equation}
    p_1(Z) = (k^2+4) \, y^2 \,
\end{equation}
where $y$ is $c_1 (\mathcal{O}(-1))$, which restricts to $x$ on the boundary.\footnote{This can be derived from the stable splitting of the tangent bundle into a direct sum of complex line bundles: $TZ \oplus \underline{\mathbb{C}} \simeq \mathcal{O}(-k) \oplus \mathcal{O}(1)\oplus \mathcal{O}(1)\oplus \mathcal{O}(1)\oplus \mathcal{O}(1)$.} Similarly, extending the gauge bundle data one finds
\begin{equation}
\widetilde{\mathcal{Q}}(j)[L^7_k(m)] = -\frac{j m^2\big(k^2 + jm^2 - 2(1+2l)\big)}{2k} \,,
\label{eq:explicitQR}
\end{equation}
in terms of the bundle data $m$, coupling $j$, and parameter specifying the lift $l$.

We can use the same formula for generators of the form $L^3_k (m) \times \text{K3}$, which physically encodes the mixed gauge gravitational anomaly.

As already discussed the discrete gauge bundle is trivial on the simply-connected K3 and only lives on $L^3_k(m)$. The gauge background $\check{C}$ is then given by
\begin{equation} \label{eq:clens3}
\check{C} = \check{c} \ast \check{c} = \big(0 \,, A_c \cup \beta(A_c) \,, 0 \big) \,,
\end{equation}
since for dimensional reasons also the characteristic class vanishes. The character is then both flat and topologically trivial. This means that the first term in \eqref{eq:QRexpform} vanishes. To compute the second term we use\footnote{In general Pontryagin classes satisfy the Whitney sum formula only up to 2-torsion. However, it can be shown that for $p_1$ the only deviation is proportional to the first Stiefel-Whitney class and thus vanishes for oriented and Spin manifolds \cite{f8ff57b0-373e-33dc-9587-d69dfe585b51}.}
\begin{equation}
    p_1 \big(L^3_k \times \text{K3} \big) = p_1\big(L^3_k \big) + p_1 (\text{K3}) \,,
\end{equation}
and find
\begin{equation}
    \widetilde{\mathcal{Q}}(j)\big[L^3_k(m) \times \text{K3} \big] = - \frac{(1+2l)j}{4} \int_{\text{K3}} p_1 \int_{L^3_k} A_c \cup \beta(A_c) = - \frac{12 (1+2l) j m^2}{k} \,.
    \label{eq:Qsecondgen}
\end{equation}
In particular, due to the prefactor $12$ we see that this is trivial, meaning integer, for $k \in \{ 2,3,4,6,12\}$, independently of the choice of $j$ or $l$. This precisely represents the fact that in these cases $L^3_k \times \text{K3}$ is not an independent generator of the bordism group.

Since the reduced quadratic refinement $\widetilde{\mathcal{Q}}$ is a bordism invariant this specifies its values on all 7-manifolds $X$, with the value determined by the bordism relations between $X$ and the generators of $\Omega^{\text{Spin}}_7 (B\mathbb{Z}_k)$. 

\vspace{0.3cm}

We now move to the derivation of anomaly constraints that are independent of $l$ and can be derived purely in terms of the non-perturbative fermion anomaly without the explicit expression above.

\section{Constraints on spectra}
\label{sec:const}

In this section we work out consistency constraints on the charged fermion spectra imposed by discrete Green-Schwarz anomaly cancellation. These restrictions arise because the non-perturbative fermion anomalies need to be canceled by a quadratic refinement of a background $\check{C}$ which only depends on the discrete gauge data. Since a quadratic refinement satisfies \eqref{eq:QRprop}, the same should be true for the fermion anomaly. These constraints are implemented by demanding anomalies to be canceled for various gauge backgrounds $N_c = m x$, affecting the background $\check{C}$.  We assume, here, that the fermion anomaly $\mathcal{A}^F$ is non-trivial. An alternative possibility is that the fermion anomaly vanishes and $j = 0$.

We will first derive consistency conditions that are independent of the choice of a specific quadratic refinement and only rely on recursion relations derived from \eqref{eq:QRprop} on seven-dimensional lens spaces. Then we determine the restrictions after fixing the quadratic refinement of the form \eqref{eq:QRexpform} leading to more detailed constraints. These are then generalized to a potential second bordism generator of type $L^3_k \times \text{K3}$.

\subsection{Universal constraints from \texorpdfstring{$L^7_k$}{L7}}
\label{subsec:constindep7}

In this section we will not fix the quadratic refinement, but rather focus on the general property \eqref{eq:QRprop}. Moreover, since the background $\check{C}$ on $L^7_k(m)$ is fully specified in terms of an element in $H^4(L^7_k;\mathbb{Z}) = \mathbb{Z}_k$, derived from the gauge field, we will keep track of it with an integer mod $k$. From \eqref{eq:QRprop} one has
\begin{equation}
    \widetilde{\mathcal{Q}}(2) = \widetilde{\mathcal{Q}}(1+1) = 2 \widetilde{\mathcal{Q}}(1) + (1,1) \,.
\end{equation}
Applying this recursively to $\widetilde{\mathcal{Q}}(n)$ one finds that all values of the quadratic refinement are determined in terms of $\widetilde{\mathcal{Q}}(1)$ and the pairing, see also \cite{Hsieh:2020jpj}, 
\begin{equation}
    \widetilde{\mathcal{Q}}(n) = n \widetilde{\mathcal{Q}}(1) + \sum_{j=1}^{n-1} (j,1) = n \widetilde{\mathcal{Q}}(1) + \Big( \frac{n(n-1)}{2}, 1\Big) \,.
    \label{eq:const1}
\end{equation}
Further taking into account the coupling constant $j$, see Section \ref{subsec:discGSlens}, this recursion relation takes the form
\begin{equation}
    \widetilde{\mathcal{Q}}(jn) = n \widetilde{\mathcal{Q}}(j) + \Big( \frac{j^2 n(n-1)}{2}, 1\Big) \,.
\end{equation}
Since the reduced quadratic refinement is trivial for vanishing gauge background, we have the additional constraint
\begin{equation}
    \widetilde{\mathcal{Q}}(k) = \widetilde{\mathcal{Q}}(0) = 0 \,,
    \label{eq:const2}
\end{equation}
since the background class lives in $\mathbb{Z}_k$.

For the seven-dimensional lens spaces $(.\,,.)$ refers to the torsion pairing, i.e., $(1,1)= - \tfrac{1}{k}$, leading to
\begin{equation}
    \widetilde{\mathcal{Q}} (jn) = n \widetilde{\mathcal{Q}} (j) - \frac{j^2 n (n-1)}{2k} \,,
    \label{eq:constj}
\end{equation}
and combining this with \eqref{eq:const2} we find
\begin{equation}
    k \widetilde{\mathcal{Q}}(j) - \frac{j^2 k (k-1)}{2k} = 0 \quad \Longrightarrow \quad \widetilde{\mathcal{Q}} (j) = \frac{j^2 (k-1)}{2k} + \frac{r}{k} \,,
\end{equation}
understood as equations mod 1. The integer $r$ parametrizes the ambiguity generated by dividing by $k$, and can be identified with the choice of the quadratic refinement. By a redefinition of $r$ this can be simplified to
\begin{equation}
    \widetilde{\mathcal{Q}} (j) = \begin{cases} \frac{r}{k} \,, \quad  \text{ for } k \text{ odd, or both } j \text{ and } k \text{ even} \,, \\
    \frac{1+2r}{2k} \,, \quad \text{ for } k \text{ even, and } j \text{ odd} \,.\end{cases}
    \label{eq:QRvalue}
\end{equation}
By evaluating $\widetilde{\mathcal{Q}}(j)$ we can therefore find the map between $r$ and $l$ in the definition \eqref{eq:explicitQR}. With this we can give general restrictions on the fermion spectrum by the cancellation condition on the non-perturbative anomalies. In particular, one needs:
\begin{equation}
    \mathcal{A}^F [L^7_k (m)] = \kappa \, \widetilde{\mathcal{Q}} (j m^2) \,,
    \label{eq:consteq}
\end{equation}
which using \eqref{eq:constj} leads to
\begin{equation}
    \mathcal{A}^F [L^7_k (m)] = \mathcal{A}^F [L^7_k (1)] -  \frac{\kappa j^2 m^2 (m^2 -1)}{2k} \,.
    \label{eq:constferm}
\end{equation}
Note that this formula is independent of the choice of the quadratic refinement and thus provides general consistency constraints. It does, however, depend on the chirality $\kappa$ of the tensor field and the discrete coupling $j$.

\subsubsection{\texorpdfstring{$\mathbb{Z}_2$}{k = 2}}

For $k =2$ the situation is fairly simple. There is only one non-trivial reduced $\eta$-invariant to consider
\begin{equation}
    \widetilde{\eta}^{\text{D}}_1 [L^7_2 (1)] = \tfrac{1}{16} \,.
\end{equation}
Since $m$ and $j$ are only meaningful modulo $2$ the only non-trivial options are $m = j =1$ and the discrete fermion anomaly reads
\begin{equation}
    \mathcal{A}^F[L^7_2 (1)] = \tfrac{1}{16} \, n_1 \,,
\end{equation}
with multiplicity $n_1$. Imposing \eqref{eq:QRvalue} one has
\begin{equation}
    n_1 = 4 \kappa (1+2r) \enspace \text{ mod } \, 16 \,,
\end{equation}
which implies
\begin{equation}
    n_1 = 4 \enspace \text{ mod } \, 8 \,,
    \label{eq:Z2const}
\end{equation}
irrespective of $\kappa$ and $r$.

\subsubsection{\texorpdfstring{$\mathbb{Z}_3$}{k = 3}}

For $k = 3$ there are two possible non-trivial charges, whose $\eta$-invariants coincide:
\begin{equation}
    \widetilde{\eta}^{\text{D}}_1 [L^7_3 (1)] = \widetilde{\eta}^{\text{D}}_2 [L^7_3 (1)] = \tfrac{1}{9} \,,
\end{equation}
from which we can evaluate the fermion anomaly in both gauge backgrounds
\begin{equation}
    \mathcal{A}^F [L^7_3 (1)] = \mathcal{A}^F [L^7_3 (2)] = \tfrac{1}{9} (n_1 + n_2) \,.
\end{equation}
The consistency condition \eqref{eq:constferm} reads
\begin{equation}
    \mathcal{A}^F [L^7_3(2)] = 4 \mathcal{A}^F[L^7_3 (1)] - \frac{12 \kappa j^2}{6} = 4 \mathcal{A}^F[L^7_3 (1)] \,,
\end{equation}
and one finds the universal condition
\begin{equation}
    n_1 + n_2 = 0 \enspace \text{ mod } \, 3 \,.
    \label{eq:Z3const}
\end{equation}
This already satisfies \eqref{eq:QRvalue} and further constraints require a specific choice of quadratic refinement.

\subsubsection{\texorpdfstring{$\mathbb{Z}_4$}{k = 4}}

The relevant $\eta$-invariants are given by
\begin{equation}
\begin{split}
    \widetilde{\eta}^{\text{D}}_1 [L^7_3 (1)] &= \widetilde{\eta}^{\text{D}}_3 [L^7_3 (1)] = \tfrac{5}{32} \,, \quad \widetilde{\eta}^{\text{D}}_2 [L^7_3 (1)] = \tfrac{1}{4} \,, \\ 
    \widetilde{\eta}^{\text{D}}_1 [\widetilde{L}^7_3 (1)] &= \widetilde{\eta}^{\text{D}}_3 [\widetilde{L}^7_3 (1)] = - \tfrac{3}{32} \,, \quad \widetilde{\eta}^{\text{D}}_2 [\widetilde{L}^7_3 (1)] = - \tfrac{1}{4} \,, 
\end{split}
\end{equation}
leading to fermion anomalies in the various gauge backgrounds
\begin{equation}
\begin{split}
    \mathcal{A}^F[L^7_4(1)] &= \mathcal{A}^F[L^7_4(3)] = \tfrac{5}{32} (n_1 + n_3) + \tfrac{1}{4} n_2 \,, \quad \mathcal{A}^F[L^7_4(2)] = \tfrac{1}{4}(n_1+n_3) \,, \\
    \mathcal{A}^F [\widetilde{L}^7_4(1)] &= - \tfrac{3}{32} (n_1 + n_3) - \tfrac{1}{4} n_2 \,, \quad \mathcal{A}^F [\widetilde{L}^7_4(2)] = - \tfrac{1}{4} (n_1 + n_3)\,.
\end{split}
\end{equation}
Implementing the recurrence relation one has
\begin{equation}
    \mathcal{A}^F [L^7_4(2)] = 4 \mathcal{A}^F [L^7_4(1)] - \frac{3 \kappa j^2 }{2} \,,
\label{eq:Z4QRconst}
\end{equation}
leading to no restrictions on $n_2$ but imposing
\begin{equation}
    n_1 + n_3 = - 4 \kappa j^2 \, \text{ mod } \, 8 = 4j \, \text{ mod } \, 8 \,.
    \label{eq:Z4const}
\end{equation}
Neither the recursion for $\mathcal{A}^F [L^7_4(3)]$ nor the conditions $\mathcal{A}^F[L^7_4(4)] = 0$ impose further restrictions. However, in this case there is another generator, the lens space with a different choice of Spin structure $\widetilde{L}^7_4(1)$. While the equivalent equation to \eqref{eq:Z4QRconst} does not lead to an extra condition, there will be another independent condition from changing the Spin structure as discussed in Section \ref{sec:spin}, constraining $n_2$ to be even.

\subsubsection{\texorpdfstring{$\mathbb{Z}_5$}{k = 5}}

The relevant $\eta$-invariants are
\begin{equation}
    \widetilde{\eta}^{\text{D}}_1 [L^7_5 (1)] = \widetilde{\eta}^{\text{D}}_4 [L^7_5 (1)] = \tfrac{1}{5} \,, \quad \widetilde{\eta}^{\text{D}}_2 [L^7_5 (1)] = \widetilde{\eta}^{\text{D}}_3 [L^7_5 (1)] = \tfrac{2}{5} \,,
\end{equation}
and the non-perturbative fermion anomalies read
\begin{equation}
\begin{split}
    \mathcal{A}^F[L^7_5(1)] &= \mathcal{A}^F[L^7_5(4)] = \tfrac{1}{5} (n_1 + n_4) + \tfrac{2}{5} (n_2 + n_3) \,, \\
    \mathcal{A}^F[L^7_5(2)] &= \mathcal{A}^F[L^7_5(3)] = \tfrac{2}{5} (n_1 + n_4) + \tfrac{1}{5} (n_2 + n_3) \,.
\end{split}
\end{equation}
The recurrence relations for $\mathcal{A}^F[L^7_5(2)]$ and $\mathcal{A}^F[L^7_5(3)]$ lead to identical constraints which is given by
\begin{equation}\label{eq:z5constr}
    n_1 + n_2 +n_3+n_4 = 3 \kappa j^2 \enspace \text{ mod } \, 5 \,.
\end{equation}
The condition $\mathcal{A}^F[L^7_5(5)] = 0$ is trivially satisfied. Constraints from the second generator will be discussed below.

\subsubsection{\texorpdfstring{$\mathbb{Z}_6$}{k = 6}}

Since $6 = 2 \cdot 3$ we expect the anomaly constraints to be a combination of those for $k = 3$ and $k = 2$ encoded in the single generator $L^7_6(1)$. The reduced $\eta$-invariants are
\begin{equation}
    \widetilde{\eta}^{\text{D}}_1 [L^7_6 (1)] = \widetilde{\eta}^{\text{D}}_5 [L^7_6 (1)] = \tfrac{35}{144} \,, \quad \widetilde{\eta}^{\text{D}}_2 [L^7_6 (1)] = \widetilde{\eta}^{\text{D}}_4 [L^7_6 (1)] = \tfrac{5}{9} \,, \quad \widetilde{\eta}^{\text{D}}_3 [L^7_6 (1)] = \tfrac{11}{16} \,.
\end{equation}
Applying the recursion relation to the fermion anomaly we extract the consistency constraint
\begin{equation}
    \mathcal{A}^F[L^7_6 (2)] = 4 \mathcal{A}^F [L^7_6(1)] - \kappa j^2 \frac{12}{12} = 4 \mathcal{A}^F [L^7_6(1)] \,,
\end{equation}
which can be rewritten in terms of the fermion spectrum as
\begin{equation}
    5 (n_1 + n_5) + 8 (n_2 + n_4) + 9 n_3 = 0 \enspace \text{ mod } \, 12 \,.
\end{equation}
Taking this equation modulo $4$ and $3$ one obtains the equivalent constraints
\begin{equation}
    \begin{split}
        n_1 + n_3 +n_5 &= 0 \enspace \text{ mod } 4 \,, \\ 
        n_1 + n_2+n_4 + n_5 &=  0 \enspace \text{ mod } 3 \,,
    \end{split}
    \label{eq:Z6const23}
\end{equation}
with an extra condition coming from $\mathcal{A}^F[L^7_6(6)] = 0$:
\begin{equation}
    11 (n_1+n_5) + 8 (n_2 + n_4) + 3 n_3 = 12 \enspace \text{ mod } \, 24
\end{equation}
Taking this equation mod 8 leads to 
\begin{equation}
    n_1+n_3+n_5 = 4 \enspace \text{ mod } 8 \,,
    \label{eq:Z6constrefinded}
\end{equation}
which is strictly stronger than \eqref{eq:Z6const23}. Indeed, the second equation of \eqref{eq:Z6const23} together with \eqref{eq:Z6constrefinded} reproduce the anomaly conditions for $\mathbb{Z}_3$ and $\mathbb{Z}_2$, respectively. This is expected on the level of bordism groups which can be determined one prime at a time, see \cite{Debray:2023yrs}, i.e.,
\begin{equation}
    \Omega^{\text{Spin}}_7 (B \mathbb{Z}_6) = \Omega^{\text{Spin}}_7 (B \mathbb{Z}_2) \oplus \Omega^{\text{Spin}}_7 (B \mathbb{Z}_3) \,,
\end{equation}
in dimension seven.\footnote{In general dimension one has to use reduced bordism groups $\widetilde{\Omega}^{\text{Spin}}_d (B \Z_k)$ with $\Omega^{\text{Spin}}_d (\text{pt})$ subtracted.}

\subsubsection{\texorpdfstring{$\mathbb{Z}_7$}{k = 7}}

With the $\eta$-invariants
\begin{equation}
    \widetilde{\eta}^{\text{D}}_1 [L^7_7(1)] = \widetilde{\eta}^{\text{D}}_6 [L^7_7(1)] = \tfrac{2}{7} \,, \quad \widetilde{\eta}^{\text{D}}_2 [L^7_7(1)] = \widetilde{\eta}^{\text{D}}_5 [L^7_7(1)] = \tfrac{5}{7} \,, \quad \widetilde{\eta}^{\text{D}}_3 [L^7_7(1)] = \widetilde{\eta}^{\text{D}}_4 [L^7_7(1)] = 0 \,,
\end{equation}
and the fermion anomalies
\begin{equation} \label{eq:Z7ferm}
    \begin{aligned}
        \mathcal{A}^F[L^7_7(1)] = \mathcal{A}^F[L^7_7(6)] &= \tfrac{2}{7}(n_1 + n_6) + \tfrac{5}{7}(n_2 + n_5) \, ,\\
        \mathcal{A}^F[L^7_7(2)] = \mathcal{A}^F[L^7_7(5)] &= \tfrac{5}{7}(n_1 + n_6) + \tfrac{2}{7}(n_3 + n_4) \, ,\\
        \mathcal{A}^F[L^7_7(3)] = \mathcal{A}^F[L^7_7(4)] &= \tfrac{2}{7}(n_2 + n_5) + \tfrac{5}{7}(n_3 + n_4) \, .
    \end{aligned}
\end{equation}
the constraints from the recursion relation are
\begin{equation}
    \mathcal{A}^F[L^7_7(2)] = 4 \mathcal{A}^F[L^7_7(1)] - \frac{6 \kappa j^2}{7} \,.
\end{equation}
In terms of the fermion spectrum this is given by
\begin{equation} \label{eq:z7constr}
   4 (n_1 + n_6) +  n_2 + n_5 + 2 (n_3 + n_4) =  \kappa j^2 \enspace \text{ mod } \, 7 \,, 
\end{equation}
with no further constraints coming from the other relations including $\mathcal{A}^F[L^7_7 (7)] = 0$.

\vspace{0.3cm}

These universal constraints can be refined by choosing a particular quadratic refinement, as we will discuss next.

\subsection{Constraints for chosen quadratic refinement}
\label{subsec:constspec}

Choosing a particular quadratic refinement is equivalent to fixing $r$ mod $k$ (or equivalently $l$ mod $k$ in the description of Section \ref{subsec:discGSlens}), which in turn fixes the value of $\widetilde{\mathcal{Q}}(j)$. This leads to refined consistency conditions, which we summarize in the following. For a parametrization of the quadratic refinement given in \eqref{eq:QRvalue} we find the following more specific constraints dependent on $r$:
\begin{equation}
    \begin{array}{c || c |}
    G & \text{anomaly cancellation depending on } r \\ \hline \hline 
    \mathbb{Z}_2 & n_1 = \kappa (4 + 8 r) \, \text{ mod } \, 16 \\ \hline
    \mathbb{Z}_3 & n_1 + n_2 = 3 \kappa r \, \text{ mod } \, 9 \\ \hline
    \mathbb{Z}_4 & 5(n_1+n_3)+8n_2 = \begin{cases}
        \kappa(4 + 8r) \, \text{ mod }\, 32   &\text{ for $j$ odd}\\
        8 \kappa r \, \text{ mod }\, 32 &\text{ for $j$ even}
    \end{cases} \\
    \hline
    \mathbb{Z}_5 & n_1 + n_4 + 2(n_2 + n_3) = \kappa r \text{ mod } \, 5 \\ \hline
    \mathbb{Z}_6 & 35 (n_1 + n_5) + 80 (n_2 +n_4) + 99 n_3 =\begin{cases}
        \kappa (12 + 24r) \, \text{ mod }\, 144   &\text{ for $j$ odd}\\
        24 \kappa r \, \text{ mod }\, 144 &\text{ for $j$ even}
    \end{cases} \\ \hline
    \mathbb{Z}_7 & 2 (n_1+n_6) + 5 (n_2 + n_5) = \kappa r \, \text{ mod } \, 7 \\ \hline
    \end{array}
\end{equation}
We further discuss the anomaly constraints in the cases $k \in \{10, 12\}$ in Appendix \ref{app:higherk}. We now move to a discussion of constraints derived from $L^3_k \times \text{K3}$.

\subsection{Constraints from \texorpdfstring{$L^3_k \times \text{K3}$}{L3 x K3} for chosen quadratic refinement}
\label{subsec:constindep3}

For $k = 5$ and $k =7$ the discussion of the bordism groups shows that there is a second, independent generator of a $\mathbb{Z}_k$ summand given by $L^3_k(m) \times \text{K3}$. We therefore expect more consistency constraints from the cancellation of anomalies on this 7-manifold. At the same time for $k \in \{2, 3, 4, 6 \}$ this does not describe an independent generator, and we expect no new constraints. These expectations are satisfied as we will now show.

Since K3 is simply-connected the discrete gauge bundle is trivial and the bundle of the product $L^3_k \times \text{K3}$ only arises due to the $L^3_k$ factor, where it is specified by an integer $m$ modulo $k$ as before. The gauge background $\check{C}$ is pulled back from the lens space factor and for dimensional reasons is both flat and topologically trivial. Its evaluation in the parameterization \eqref{eq:QRexpform} was explicitly discussed in Section \ref{subsec:discGSlens}. The fermion anomaly can be evaluated by using the formula \eqref{eq:etaprod} that evaluates the $\eta$-invariant on product manifolds. 

\subsubsection{No independent constraints for \texorpdfstring{$k \in \{2,3,4,6,12\}$}{k = 2,3,4,6,12}}

First, we demonstrate that anomaly cancellation on $L^3_k \times \text{K3}$ does not lead to independent constraints for $k \in \{ 2,3,4,6,12\}$, for which the product is bordant to a number of seven-dimensional lens spaces. Even more restrictively, since the quadratic refinement contribution is trivial \eqref{eq:Qsecondgen} for any $j,l$, the fermion anomaly needs to vanish as well.

For $\mathbb{Z}_2$ we have the fermion anomaly
\begin{equation}
    \widetilde{\eta}^{\text{D}}_1 [L^3_2(1) \times \text{K3}] = \tfrac{1}{2} \quad \Longrightarrow \quad \mathcal{A}^F [L^3_2(1) \times \text{K3}] = \tfrac{1}{2} n_1 \,,
\end{equation}
which is trivial once we impose \eqref{eq:Z2const}. Thus, the constraints are contained in those from $L^7_2(1)$. 

For $\mathbb{Z}_3$ one finds
\begin{equation}
    \mathcal{A}^F [L^3_3(1) \times \text{K3}] = 
    \mathcal{A}^F [L^3_3(2) \times \text{K3}] = \tfrac{1}{3} (n_1 + n_2) \,,
\end{equation}
which becomes trivial after using \eqref{eq:Z3const}.

The fermion anomalies in the case of $\mathbb{Z}_4$ gauge group are given by
\begin{equation}
    \mathcal{A}^F[L^3_4(1) \times \text{K3}] = \mathcal{A}^F[L^3_4(3) \times \text{K3}] = \tfrac{1}{4} (n_1 + n_3) \,, \quad \mathcal{A}^F[L^3_4(2) \times \text{K3}] = 0 \,,
\end{equation}
trivial for all spectra consistent with \eqref{eq:Z4const}.

Again, we expect that for $k = 6$ follow from imposing the constraints for $k =2$ and $k =3$ simultaneously. The fermion anomalies in the different backgrounds are given by
\begin{equation}
    \begin{split}
    \mathcal{A}^F[L^3_6(1) \times \text{K3}] &= \tfrac{2}{3} (n_1 + n_2 + n_4 +n_5) + \tfrac{1}{2} (n_1 + n_3 + n_5) \,, \\
    \mathcal{A}^F[L^3_6(2) \times \text{K3}] &= \tfrac{2}{3} (n_1 + n_2 + n_4 +n_5)\,, \\
    \mathcal{A}^F[L^3_6(3) \times \text{K3}] &= \tfrac{1}{2} (n_1 + n_3 + n_5) \,, 
    \end{split}
\end{equation}
all of which are trivialized by the conditions \eqref{eq:Z6const23}.

In Appendix \ref{app:higherk} we perform the same computation for $\Z_{12}$, and show the fermion anomaly is also trivialized.

\subsubsection{\texorpdfstring{$\mathbb{Z}_5$}{k = 5}}

The reduced $\eta$-invariants for $\mathbb{Z}_5$ on the second generator are given by
\begin{equation}
    \widetilde{\eta}^{\text{D}}_1 [L^3_5(1) \times \text{K3}] = \widetilde{\eta}^{\text{D}}_4 [L^3_5(1) \times \text{K3}] = \tfrac{1}{5} \,, \quad \widetilde{\eta}^{\text{D}}_2 [L^3_5(1) \times \text{K3}] = \widetilde{\eta}^{\text{D}}_3 [L^3_5(1) \times \text{K3}] = \tfrac{4}{5} \,,
\end{equation}
leading to the fermion anomalies
\begin{equation}
\begin{split}
    \mathcal{A}^F [L^3_5(1) \times \text{K3}] &= \mathcal{A}^F [L^3_5(4) \times \text{K3}] = \tfrac{1}{5}(n_1+n_4) + \tfrac{4}{5}(n_2+n_3) \,, \\
    \mathcal{A}^F [L^3_5(2) \times \text{K3}] &= \mathcal{A}^F [L^3_5(3) \times \text{K3}] = \tfrac{4}{5}(n_1+n_4) + \tfrac{1}{5}(n_2+n_3) \,.
\end{split}
\end{equation}
They are all proportional to each other, so to cancel anomalies it is enough to impose $\mathcal{A}^F = \widetilde{\QR}(j)$ on $L^3_5(1) \times \text{K3}$, and doing so leads to
\begin{equation}
    n_1 + n_4 +4(n_2+n_3) = \kappa j(3+l) \,.
\end{equation}
By imposing the constraint \eqref{eq:z5constr} we found on the other generator, we find we can completely fix the spectrum (mod 5) as
\begin{equation}\label{eq:spectraZ5}
    \begin{aligned}
        n_1+n_4 & = 3\kappa j l -\kappa j(j+1) \hspace{4mm} \text{ mod 5} \,, \\
        n_2+n_3 & = 2 \kappa j l - \kappa j (j-1)  \hspace{4mm} \text{ mod 5} \,.
    \end{aligned}
\end{equation}
Since the spectra are fixed as a function of $\kappa,l,j$, so are the fermion anomalies. As a sanity check of our procedure, we can check that anomaly cancellation is still happening on the other generator $L_5^7(1)$. What we find is
\begin{equation}
    \begin{aligned}
        \mathcal{A}^F[L_5^7(1)] &= \frac{\kappa j}{5}(1+2j+2l) \,, \\
    \kappa \widetilde{\QR}(j)[L_5^7(1)] &= -\frac{\kappa j}{10}(3+j-4l) \,,
    \end{aligned}
\end{equation}
which can be shown to coincide up to integers as functions of $\kappa,l,j$.

\subsubsection{\texorpdfstring{$\mathbb{Z}_7$}{k = 7}}

The reduced $\eta$-invariants for $\mathbb{Z}_7$ on the second generator are given by
\begin{equation}
    \begin{aligned}
        \widetilde{\eta}_1^\text{D}[L_7^3(1) \times \text{K3}] =  \widetilde{\eta}_6^\text{D}[L_7^3(1) \times \text{K3}]& = \tfrac{1}{7}, \qquad \widetilde{\eta}_2^\text{D}[L_7^3(1) \times \text{K3}] =  \widetilde{\eta}_5^\text{D}[L_7^3(1) \times \text{K3}] = \tfrac{4}{7} \,, \\
        \widetilde{\eta}_3^\text{D}[L_7^3(1) \times \text{K3}] & =  \widetilde{\eta}_4^\text{D}[L_7^3(1) \times \text{K3}] = \tfrac{2}{7} \,,
    \end{aligned}
\end{equation}
leading to the fermion anomalies
\begin{equation}
    \begin{aligned}
        \mathcal{A}^F[L_7^3(1) \times \text{K3}] = \mathcal{A}^F[L_7^3(6) \times \text{K3}] &= \tfrac{1}{7}(n_1 + n_6) + \tfrac{4}{7}(n_2 + n_5) + \tfrac{2}{7}(n_3 + n_4) \,,\\
        \mathcal{A}^F[L_7^3(2) \times \text{K3}] = \mathcal{A}^F[L_7^3(5) \times \text{K3}] &= \tfrac{4}{7}(n_1 + n_6) + \tfrac{2}{7}(n_2 + n_5) + \tfrac{1}{7}(n_3 + n_4) \,,\\
        \mathcal{A}^F[L_7^3(3) \times \text{K3}] = \mathcal{A}^F[L_7^3(4) \times \text{K3}] &= \tfrac{2}{7}(n_1 + n_6) + \tfrac{1}{7}(n_2 + n_5) + \tfrac{4}{7}(n_3 + n_4) \,.
    \end{aligned}
\end{equation}
Again, they are all proportional to each other, so to cancel anomalies it is enough to impose $\mathcal{A}^F = \widetilde{\QR}(j)$ on $L^3_7(1) \times \text{K3}$, and doing so leads to
\begin{equation}
    n_1+n_6 + 4(n_2+n_5) + 2 (n_3+n_4) = 2 j \kappa (1+2l)  \hspace{4mm} \text{ mod 7}
\end{equation}
This, together with \eqref{eq:z7constr} does not fix the whole spectrum, but it fixes the combinations
\begin{equation}\label{eq:z7spectra}
    \begin{aligned}
         n_3+n_4-(n_2+n_5) &= j \kappa (2 + 4l + 3j)  \hspace{4mm} \text{ mod 7} \\
         n_2+n_5-(n_1+n_6) &= j \kappa(3 - l + 2j) \hspace{4mm} \text{ mod 7}
    \end{aligned}
\end{equation}
which appear in the fermion anomaly on the other generator. Then we again need to check anomaly cancellation is still happening, and find
\begin{equation}
    \begin{aligned}
        \mathcal{A}^F[L_7^7(1)] &= \frac{\kappa j}{7}(1+2l+3j) \,, \\
         \widetilde{\QR}(j)[L_7^7(1)] &= -\frac{\kappa j}{14}(5+j-4l) \,, 
    \end{aligned}
\end{equation}
which are equivalent modulo integers for all $\kappa,j,l$.

For higher values of $k$ the spectra have enough degrees of freedom $n_q$ that the fermion anomalies will not be fixed by recursion and cancellation on only one generator, so there is no need to perform this check.

\section{Consistency under change of Spin structure}
\label{sec:spin}

In this section we wish to provide another consistency check of our procedure, which has to do with the behavior of the differential lift of the Wu class under a change of Spin structure. On Spin manifolds a natural choice \cite{Hopkins:2002rd,Hsieh:2020jpj} for this lift is given by \eqref{eq:wu4}, but we have already seen that it is not unique, for example by choosing an integral class such as \eqref{eq:genWu}.

These lifts have a well-defined behavior under a change of Spin structure of the underlying manifold, and the $\eta$-invariants representing fermion anomalies also naturally depend on the choice of Spin structure. Therefore we need to check that these two transformations on either side of \eqref{eq:consteq} match. 

The set of Spin structures on a manifold $X$ is a torsor over the group $H^1(X;\Z_2)$, which means we can represent a change of Spin structure by a class $\alpha \in H^1(X;\Z_2)$. The transformation properties of a differential lift $\check{\nu}$ of the Wu class under a change of Spin structure which we denote by $s \mapsto s + \alpha$ is computed in Appendix E of \cite{Hopkins:2002rd}, and reads
\begin{equation} \label{eq:nushift}
        \check{\nu}(s+ \alpha) = \check{\nu}(s) + \delta \Big( \nu(s) \sum_{k \geq 1} \alpha^{2^k-1} \Big) \,.
\end{equation}
In this formula, the quantities without a check denote ordinary cohomology classes, and in particular $\nu(s) = \sum_i \nu_i(s) \in H^*(X;\Z)$ is the total integral lift of the Wu class associated to $s$. The implicit product is given by the cup product degree by degree, and the map $\delta$ is the composite operation
\begin{equation} \label{eq:bockspin}
        H^p(X;\Z_2) \longrightarrow H^p(X;U(1)) \longrightarrow \check{H}^{p+1}(X) \,,
\end{equation}
where the first arrow is induced by the usual inclusion $\Z_2 \hookrightarrow U(1)$ in the coefficients, and the second arrow is the Bockstein homomorphism that describes flat characters described in footnote \ref{footnote:flatchar}. We are interested in the degree 4 component of $\check{\nu}$, for which the shift is simply given by $\delta(\alpha^3)$. 

The change of the quadratic refinement inducing a change in the anomaly cancellation $\mathcal{A}^B_{\text{GS}}$ is given on the bordism generators by:
\begin{equation} \label{table:spinT}
    \begin{array}{c || c | c |}
    \Delta \mathcal{A}^B_\text{GS}[L_k^7(m)] & k = 4\ell +2 &  k = 4
\ell  \\ \hline \hline
    m \text{  odd} & \frac{1}{2}& 0 \\ \hline
     m \text{  even} & 0 & 0 \\ \hline
    \end{array}
    \qquad
    \begin{array}{c || c | c |}
    \Delta \mathcal{A}^B_\text{GS}[L_k^3(m) \times \text{K3}] & k = 4\ell +2 &  k = 4\ell  \\ \hline \hline
    m \text{  odd} & 0 & 0 \\ \hline
    m \text{  even} & 0 & 0 \\ \hline
    \end{array}
\end{equation}
For a detailed proof of these statements see Appendix \ref{app:spinQR}. From the tables it is clear what we should require of the fermion anomalies: when $k = 4\ell+2$ the anomaly in an odd $m$ background should shift by $1/2$ under a change of Spin structure, while in all other cases it should be invariant.

Let us now compute this variation for the fermion anomalies as a function of the spectrum, by studying the dependence of the $\eta$-invariants on the Spin structure. From \eqref{eq:etaspin} we see that we can view the effect of the change of Spin structure on $L_k^n$ on $\eta$-invariants as an effective shift in the charge of all fermions. Then the change in the pure $\Z_k$ gauge anomaly of a single fermion of charge $q$ is
\begin{equation}\label{eq:fermspin}
    \Delta \widetilde{\eta}^{\text{D}}_q[L_k^n (m)] = \big(\eta^{\text{D}}_{q+\frac{k}{2}}[L_k^n (m)]- \eta^{\text{D}}_{\frac{k}{2}}[L_k^n (m)]\big) - \big(\eta^{\text{D}}_{q}[L_k^n (m)]- \eta^{\text{D}}_{0}[L_k^n (m)]\big) \,.
\end{equation}
For the cases of interest these changes are given by:
\begin{equation}
    \begin{array}{c || c | c |}
    \Delta \widetilde{\eta}^{\text{D}}_q[L_k^7 (m)] & k = 4\ell +2 &  k = 4\ell  \\ \hline \hline
    q = 2l+1 & -\frac{2\ell + 1}{8} & \frac{\ell}{4} \\ \hline
    q = 2l & 0 & \frac{q}{4} \\ \hline
    \end{array}
    \qquad \qquad
    \begin{array}{c || c | c |}
    \Delta \widetilde{\eta}^{\text{D}}_q[L_k^3(m) \times \text{K3}] & k = 4\ell +2 &  k = 4\ell  \\ \hline \hline
    q = 2l+1 & 0 & 0 \\ \hline
    q = 2l & 0 & 0 \\ \hline
    \end{array}
\end{equation}
Therefore on the generators $L_k^3(m) \times \text{K3}$ the behavior of the tensor and fermion anomalies is automatically matched, independently of the spectrum. For the generator $L^7_k(m)$ this matching is non-trivial.

For even background $m$, the effective charges $m \cdot q$ are all even, so the fermion anomaly is invariant for all spectra, consistently with the change in the tensor anomaly \eqref{eq:tensSpin}. For odd background $m$, however, $m\cdot q = q$ mod 2, so the transformations are the same as described above. We can then write the full change as
\begin{equation}
    \Delta \mathcal{A}^F[L_k^7(m)] = \sum_q n_q \Delta \widetilde{\eta}^{\text{D}}_{mq} =
    \begin{cases}
        \scalebox{1.1}{$-\frac{2\ell+1}{8}\sum\limits_\text{$q$ odd} n_q $} &\text{for $k = 4\ell+2$, $m$ odd} \,, \\
        \scalebox{1.1}{$\frac{\ell}{4}\sum\limits_\text{$q$ odd} n_q$} + \scalebox{1.1}{$\sum\limits_{q \text{ even}} \frac{m q}{4} n_q$} &\text{for $k = 4\ell$, $m$ odd} \,, \\
        0 &\text{otherwise} \,.
    \end{cases}
\end{equation}
From which we see that to match it to \eqref{table:spinT} we require 
\begin{equation}   \label{eq:spinfinal}
    \begin{aligned}
        \sum_\text{$q$ odd} n_q = 4 \enspace \text{ mod } \, 8 \,, \quad &\text{for $k = 4\ell+2$} \,, \\
        \ell \sum_\text{$q$ odd} n_q + 2\sum_{q = 2 \text{ mod 4}}n_q  = 0 \enspace \text{ mod } \, 4 \,, \quad &\text{for $k = 4\ell$} \,.
    \end{aligned}
\end{equation}
One can check that in all the cases discussed in Section \ref{sec:const} and Appendix \ref{app:higherk} these constraints are compatible with those imposed by anomaly cancellation, and in particular for $k= 4\ell+2$ they are implied by the recursion relations. This is not surprising; for these groups the lens spaces with two different Spin structures are not independent generators of the bordism group. We can view the results of this Section as another non-trivial consistency check of our procedure. For $k = 4\ell$, however, the lens spaces with different Spin structures are independent generators of the bordism group, which is reflected in the fact that the equations in \eqref{eq:spinfinal} implement additional constraints on top of those discussed in Section \ref{sec:const}.

For example in the case $k = 4$, i.e., $\ell =1$, we obtain the independent constraint
\begin{equation}
    n_1 + n_3 + 2 n_2 = 0 \enspace \text{ mod } \, 4 \, .
\end{equation}
Since we found in \eqref{eq:Z4const} that $n_1+n_3 = 0$ mod 4, this is an independent constraint
\begin{equation}
    n_2 = 0\enspace \text{ mod } \, 2 \,,
\end{equation}
restricting the number of fermions with charge $2$.
\\There is another interpretation \footnote{We thank Miguel Montero for pointing out this possibility.} for the second constraints in \eqref{eq:spinfinal}: the results in this section assume that the quadratic refinements are those through the natural lifts \eqref{eq:genWu} of the Wu class on spin manifolds, as in \cite{Hopkins:2002rd}. Models which satisfy all other constraints but \eqref{eq:spinfinal} raise the interesting question of whether there are consistent theories with a different choice of quadratic refinement, implementing an alternative constraint on the fermion spectrum. If so, it would be very interesting to find how their UV completions differ from the quadratic refinement in \cite{Hopkins:2002rd} which was motivated by M-theory.

\section{Conclusion}
\label{sec:concl}

In this work we studied the cancellation of discrete gauge anomalies in six-dimensional theories, via a generalization of the Green-Schwarz mechanism. The chirality of the 2-form field demands a choice of a quadratic refinement that also influences the anomaly constraints. However, using universal properties of such quadratic refinements we derive strong restrictions on the spectrum of charged fermions. These are in agreement with earlier considerations in the literature and relate the anomaly free spectra to the anomaly inflow onto strings coupling to the 2-form fields. Once a specific quadratic refinement is chosen, for which we provide an explicit formula and which is part of the defining data of the theory, the anomaly constraints are refined further. Finally, we can implement consistency conditions from changes of the quadratic refinement under a change of Spin structure for $k$ even. These provide independent constraints in the case $k = 4 \ell$ and consistency checks for $k = 4 \ell +2$. Our findings further reflect the structure of generators for the associated bordism groups that classify the potential non-perturbative anomalies, as necessary.

In an upcoming paper \cite{DOT2025} our anomaly constraints are verified for many six-dimensional supergravity models constructed from F-theory with multi-sections, providing a top-down counterpart to this work. It would be very interesting to explore how this generalizes after the inclusion of more, potentially continuous, gauge groups and several tensor fields, see \cite{Monnier:2018nfs}. In the presence of several tensor fields the quadratic refinement includes the lattice pairing and include anomaly inflow on the various strings coupling to the 2-form fields, suggesting a more flexible cancellation, which can relax the constraints derived in this work. Having this generalization available, on might confront the various seemingly consistent six-dimensional supergravity models with refined anomaly constraints \cite{Kumar:2009ae, Seiberg:2011dr, Taylor:2018khc, Tarazi:2021duw, Hamada:2021yxy, Hamada:2024oap, Loges:2024vpz, Kim:2024hxe}, potentially banishing some of them to the Swampland. A more geometrical understanding of the anomaly constraints might also be a gateway towards string universality in lower dimensions \cite{Kumar:2009us, Adams:2010zy, Montero:2020icj, Cvetic:2020kuw, Bedroya:2021fbu, Hamada:2021bbz, Bedroya:2023tch}, where the six-dimensional theories with eight supercharges form a highly interesting class of models.

Six-dimensional supergravity theories with $\mathcal{N}=(1,0)$ supersymmetry also possess interesting duality symmetries acting on the tensor fields. Thus, for several 2-forms there is a possibility of an interplay between the duality symmetry and the generalized form of the Green-Schwarz cancellation, which might further provide restrictions on the allowed fermion spectra. 

Finally, one might speculate on whether there are other ways to cancel the discrete gauge anomalies. For instance by the introduction of new topological degrees of freedom, which can contribute to the gauge variation by associating their generalized symmetries with gauge backgrounds \cite{Garcia-Etxebarria:2017crf}, see also \cite{Debray:2021vob}. This form of anomaly cancellation might further demand the introduction of new dynamical objects with interesting properties under anomaly inflow.

\section*{Acknowledgments}

MD thanks Arun Debray, Paul-Konstantin Oehlmann, and Thorsten Schimannek for many valuable discussions and collaborations on related projects.
We are very grateful to Ivano Basile, Arun Debray, Miguel Montero and Paul-Konstantin Oehlmann for comments on the Draft, and want to especially thank Arun for his detailed and insightful input.  MT is supported by the FPI grant PRE2022-102286 from Spanish National Research Agency from the Ministry of Science and Innovation.

\begin{appendix}

\section{Reduced quadratic refinement}
\label{app:redQR}

Let us assume that the background $\check{C}\in\check{H}^4(X)$ entering the quadratic refinement term depends on both gauge and gravitational data of $X$. We split the background accordingly\footnote{Since we consider $X$ to be Spin, the first and second Stiefel-Whitney classes vanish, and there is no mixed term between gauge and gravitational background in dimension $4$.}
\begin{equation}
    \check{C} = \check{C}_k + \check{C}_0 \,,
\end{equation}
where $\check{C}_0$ captures the gravitational part and $\check{C}_k$ the gauge part, respectively. The reduced quadratic refinement \eqref{eq:redQ} is then given by
\begin{equation}
    \widetilde{\mathcal{Q}}(\check{C}) = \mathcal{Q}(\check{C}_k + \check{C}_0) - \mathcal{Q}(\check{C}_0) \,,
\end{equation}
identifying $\mathcal{Q}_0 = \mathcal{Q}(\check{C}_0)$. For trivial gauge field the background $\check{C}_k$ vanishes as does its reduced quadratic refinement as intended. We use the defining properties of a quadratic refinement \eqref{eq:QRprop} to show
\begin{equation}
\begin{split}
    \mathcal{Q}(\check{C}_k^{(1)} + \check{C}_k^{(2)} + \check{C}_0) &= (\check{C}_k^{(1)}, \check{C}_k^{(2)} + \check{C}_0) + \mathcal{Q} (\check{C}_k^{(1)}) + \mathcal{Q}(\check{C}_k^{(2)} + \check{C}_0) - \mathcal{Q}(0) \\
    &= (\check{C}_k^{(1)}, \check{C}_k^{(2)}) + (\check{C}_k^{(1)}, \check{C}_0 ) + \mathcal{Q} (\check{C}_k^{(1)}) + \mathcal{Q}(\check{C}_k^{(2)} + \check{C}_0) - \mathcal{Q}(0) \\
    &= (\check{C}_k^{(1)}, \check{C}_k^{(2)}) + \mathcal{Q}(\check{C}_k^{(1)} + \check{C}_0) + \mathcal{Q}(\check{C}_k^{(2)} + \check{C}_0) - \mathcal{Q}(\check{C}_0) \,.
\end{split}
\label{eq:modbackgqr}
\end{equation}
Using this we find
\begin{equation}
     \widetilde{\mathcal{Q}} (\check{C}_k^{(1)} + \check{C}_k^{(2)} + \check{C}_0) = (\check{C}_k^{(1)}, \check{C}_k^{(2)}) + \widetilde{\mathcal{Q}} (\check{C}_k^{(1)} + \check{C}_0) + \widetilde{\mathcal{Q}} (\check{C}_k^{(2)} + \check{C}_0) \,,
\end{equation}
which therefore also satisfies the defining property for a quadratic refinement of the torsion pairing $(. \,, .)$.

\section{Bordism groups for cyclic groups of prime order}
\label{app:bordprime}

In this Appendix we prove the splitting of $\Omega^{\text{Spin}}_7 (B\mathbb{Z}_k)$ for $k = p$ a prime bigger than $3$. We do so by using various homotopy equivalences that were explained in more detail in \cite{Debray:2023yrs}.

The map $\Omega^{\text{Spin}}_d \rightarrow \Omega^{\text{SO}}_d$ is a $p$-local equivalence for odd primes. At prime $p$ one finds, \cite{BP66}, that the oriented bordism groups split as sum of Brown-Peterson homology
\begin{equation}
    BP_d (\text{pt}) \cong \mathbb{Z}_{(p)} [v_1,v_2, \dots] \,,
\end{equation}
with generators $v_i$ of degree $|v_i| = 2(p^i - 1)$. The symbol $\mathbb{Z}_{(p)}$ means the ring of rational numbers whose denominators are not divisible by $p$, which we use to localize at prime $p$. Comparing this to 
\begin{equation}
    \Omega^{\text{SO}}_d(\text{pt}) \otimes \mathbb{Z}_{(p)} \cong \mathbb{Z}_{(p)} [x_4,x_8,\dots] \,,
\end{equation}
with degrees $|x_{4i}| = 4i$. With this one finds for $p$ an odd prime and $d < 13$
\begin{equation}
\begin{split}
    \Omega^{\text{SO}}_d (X) \otimes \mathbb{Z}_{(3)} &= BP_d(X) \oplus BP_{d-8} (X) \oplus BP_{d-12}(X) \,, \\
    \Omega^{\text{SO}}_d (X) \otimes \mathbb{Z}_{(5)} &= BP_d(X) \oplus BP_{d-4} (X) \oplus BP_{d-8} (X) \oplus BP_{d-12}(X) \,, \\
    \Omega^{\text{SO}}_d (X) \otimes \mathbb{Z}_{(7)} &= BP_d(X) \oplus BP_{d-4}(X) \oplus BP_{d-8}(X)^{\oplus 2} \oplus BP_{d-12}(X)^{\oplus 2} \,, \\
    \Omega^{\text{SO}}_d (X) \otimes \mathbb{Z}_{(p)} &= BP_d(X) \oplus BP_{d-4}(X) \oplus BP_{d-8}(X)^{\oplus 2} \oplus BP_{d-12}(X)^{\oplus 3} \,, \quad p > 7 \,,  
\end{split}
\end{equation}
Next, we use that $BP$ homology can be expressed in terms of $\ell$ homology \cite{BBDG89}
\begin{equation}
    BP_d(B\mathbb{Z}_p) \cong \ell_d(B\mathbb{Z}_p) \otimes \mathbb{Z}_{(p)} [v_2,v_3, \dots] \,.
\end{equation}
Finally, $\ell_d$ can be determined in terms of connective complex K-theory groups $ku$ localized at prime $p$ given by
\begin{equation}
    ku_d(X) \otimes \mathbb{Z}_{(p)} \cong \bigoplus_{j = 0}^{p-2} \ell_{d-2j} (X) \,.
\end{equation}
Moreover, we know the these groups from \cite{BG03} and they are given by
\begin{equation}
    \widetilde{ku}_d (B\mathbb{Z}_p) \otimes \mathbb{Z}_{(p)} = \begin{cases} \mathbb{Z}_{p^{j+1}}^{\oplus s} \oplus \mathbb{Z}_{p^{j}}^{\oplus p -1-s} \,, \quad \text{if } d = 2j(p-1)+2s-1 \enspace \text{with } 0 < s \leq p-1 \,, \\
    0 \,, \quad \text{otherwise}\end{cases} \,.
\end{equation}
Thus, we see that the relevant dimensions for values of $j$ are given by
\begin{equation}
    \begin{split}
        j = 0:& \quad d \in \{1, 3, \dots, 2p-3 \} \,, \\
        j = 1:& \quad d \in \{2p-1, 2p+1, \dots, 4p-5\} \,, \\
        j = 2:& \quad d \in \{4p-3, 4p-1, \dots, 6p-7\} \,, \dots
    \end{split}
\end{equation}
With this we see that for odd primes below dimension $12$ we find 
\begin{equation}
    \begin{array}{c || c | c | c | c }
    p & 3 & 5 & 7 & 11 \\ \hline
    j_{\text{max}} & 3 & 2 & 1 & 0
    \end{array}
\end{equation}
with $j_{\text{max}} = 0$ for all higher primes. The corresponding $ku$ groups are (they vanish in even degree)
\begin{equation}
    \begin{array}{c || c | c | c | c | c | c |}
    d & 1 & 3 & 5 & 7 & 9 & 11 \\ \hline \hline
    \widetilde{ku}(B\mathbb{Z}_3) \otimes \mathbb{Z}_{(3)} & \mathbb{Z}_3 & \mathbb{Z}_3^{\oplus 2} & \mathbb{Z}_9 \oplus \mathbb{Z}_3 & \mathbb{Z}_9^{\oplus 2} & \mathbb{Z}_{27} \oplus \mathbb{Z}_9 & \mathbb{Z}_{27}^{\oplus 2} \\ \hline
    \widetilde{ku}(B\mathbb{Z}_5) \otimes \mathbb{Z}_{(5)} & \mathbb{Z}_{5} &  \mathbb{Z}_{5}^{\oplus 2} & \mathbb{Z}_{5}^{\oplus 3} & \mathbb{Z}_{5}^{\oplus 4} & \mathbb{Z}_{25} \oplus \mathbb{Z}_5^{\oplus 3} & \mathbb{Z}_{25}^{\oplus 2} \oplus \mathbb{Z}_5^{\oplus 2}\\ \hline
    \widetilde{ku}(B\mathbb{Z}_7) \otimes \mathbb{Z}_{(7)} & \mathbb{Z}_{7} &  \mathbb{Z}_{7}^{\oplus 2} & \mathbb{Z}_{7}^{\oplus 3} & \mathbb{Z}_{7}^{\oplus 4} & \mathbb{Z}_{7}^{\oplus 5} & \mathbb{Z}_{7}^{\oplus 6} \\ \hline
    \widetilde{ku}(B\mathbb{Z}_p) \otimes \mathbb{Z}_{(p)} \,, p>7 & \mathbb{Z}_p &  \mathbb{Z}_p^{\oplus 2} & \mathbb{Z}_p^{\oplus 3}  & \mathbb{Z}_p^{\oplus 4}  & \mathbb{Z}_p^{\oplus 5} & \mathbb{Z}_p^{\oplus 6} \\ \hline
    \end{array}
\end{equation}
From this it is not hard to determine the reduced $\ell$ homology groups
\begin{equation}
    \begin{array}{c || c | c | c | c | c | c |}
    d & 1 & 3 & 5 & 7 & 9 & 11 \\ \hline \hline
    \widetilde{\ell}_d (B\mathbb{Z}_3) & \mathbb{Z}_3 & \mathbb{Z}_3 & \mathbb{Z}_9 & \mathbb{Z}_9 & \mathbb{Z}_{27} & \mathbb{Z}_{27} \\ \hline
    \widetilde{\ell}_d (B\mathbb{Z}_5) & \mathbb{Z}_5 & \mathbb{Z}_5 & \mathbb{Z}_5 & \mathbb{Z}_5 & \mathbb{Z}_{25} & \mathbb{Z}_{25} \\ \hline
    \widetilde{\ell}_d (B\mathbb{Z}_7) & \mathbb{Z}_7 & \mathbb{Z}_7 & \mathbb{Z}_7 & \mathbb{Z}_7 & \mathbb{Z}_7 & \mathbb{Z}_7 \\ \hline
    \widetilde{\ell}_d (B\mathbb{Z}_p) \,, p>7 & \mathbb{Z}_p & \mathbb{Z}_p & \mathbb{Z}_p & \mathbb{Z}_p & \mathbb{Z}_p & \mathbb{Z}_p\\ \hline
    \end{array}
\end{equation}
We can use these to determine the reduced Spin bordism groups, since in the range we are interested in one has $\ell_d(X) \cong BP_d(X)$ and we find for odd primes $p$
\begin{equation}
    \begin{array}{c || c | c | c | c | c | c |}
    d & 1 & 3 & 5 & 7 & 9 & 11 \\ \hline \hline
    \widetilde{\Omega}^{\text{Spin}}_{d} (B\mathbb{Z}_3) & \mathbb{Z}_3 & \mathbb{Z}_3 & \mathbb{Z}_9 & \mathbb{Z}_9 & \mathbb{Z}_{27} \oplus \mathbb{Z}_3 & \mathbb{Z}_{27} \oplus \mathbb{Z}_3\\ \hline 
    \widetilde{\Omega}^{\text{Spin}}_{d} (B\mathbb{Z}_5) & \mathbb{Z}_5 & \mathbb{Z}_5 & \mathbb{Z}_5 \oplus \mathbb{Z}_5 & \mathbb{Z}_5 \oplus \mathbb{Z}_5 & \mathbb{Z}_{25} \oplus \mathbb{Z}_5^{\oplus 2} & \mathbb{Z}_{25} \oplus \mathbb{Z}_5^{\oplus 2} \\ \hline
    \widetilde{\Omega}^{\text{Spin}}_{d} (B\mathbb{Z}_7) & \mathbb{Z}_7 & \mathbb{Z}_7 & \mathbb{Z}_7 \oplus \mathbb{Z}_7 & \mathbb{Z}_7 \oplus \mathbb{Z}_7 & \mathbb{Z}_7^{\oplus 4} & \mathbb{Z}_7^{\oplus 4} \\ \hline
    \widetilde{\Omega}^{\text{Spin}}_{d} (B\mathbb{Z}_p) \,, p>7 & \mathbb{Z}_p & \mathbb{Z}_p & \mathbb{Z}_p \oplus \mathbb{Z}_p & \mathbb{Z}_p \oplus \mathbb{Z}_p & \mathbb{Z}_p^{\oplus 4} & \mathbb{Z}_p^{\oplus 4} \\ \hline
    \end{array}
\end{equation}
Here the $\widetilde{\Omega}^{\text{Spin}}_d$ refer to reduced bordism groups without the gravitational contribution $\Omega^{\text{Spin}}_d (\text{pt})$. This shows that the extension problem splits for $p = 5$ in dimension 7 and lower and for $p>5$ in dimension 11 and lower.

\section{\texorpdfstring{$\eta$}{Eta}-invariants for lens spaces} 
\label{app:eta}

We collect here some formulae to compute $\eta$-invariants on lens spaces, for the derivation see e.g. \cite{Hsieh:2020jpj,Debray:2023yrs,Basile:2023zng}.
Consider a lens space $L_k^{2n-1}(t_1,...,t_n)$, with $t_i$ integers co-prime to $k$, defined through the quotient $S^{2n-1}/\Z_k$ acting on the ambient space $\mathbb{C}^n$ as
\begin{equation}
    z_i \mapsto \text{exp} \Big(2 \pi i  \frac{t_i}{k}\Big) z_i \,.
\end{equation}
The $\eta$-invariant associated to a Dirac fermion with charge $q$ under $\Z_k$ on such a space is given as
\begin{equation}
    \eta^{\text{D}}_q \left[ L_k^{2n-1}(t_1,...,t_n)\right] = -\frac{1}{k(2i)^n} \sum_{l=1}^{k-1} \frac{e^{-2 \pi i l q /k}}{\text{sin}(\pi l t_1 /k)...{\text{sin}(\pi l t_n /k)}} \,.
\end{equation}
The spaces we are concerned about in the main text are recovered as $t_1 = ... = t_n = 1$, for which a simpler polynomial expression is available, see also \cite{Basile:2023zng}, valid modulo integers:
\begin{equation}
    \begin{aligned}
        \eta^{\text{D}}_q \left[ L_k^{3}\right] &= \frac{k^2-6kq+6q^2-1}{12k} \,, \\
        \eta^{\text{D}}_q \left[ L_k^{7}\right] &= -\frac{k^4+ 10k^2 - 30k^2 q^2 - 60kq +60k q^3 +60 q^2-30q^4-11}{720k} \,.
    \end{aligned}
\end{equation}
Subtracting $\widetilde{\eta}_0$ yields the bordism invariants we are interested in:
\begin{equation} \label{eq:polyeta}
    \begin{aligned}
        \widetilde{\eta}^{\text{D}}_q \left[ L_k^{3}\right] &= \frac{q(q-k)}{2k} \,,\\
        \widetilde{\eta}^{\text{D}}_q \left[ L_k^{7}\right] &= \frac{k^2q^2+2kq(1-q^2)+q^4-2q^2}{24k} \,.
    \end{aligned}
\end{equation}

\subsection{Change of Spin structure}

As explained in Section \ref{sec:spin}, the quantity we need to compute to study the behavior of the fermion gauge anomaly under a change of Spin structure is \eqref{eq:fermspin}.
We are interested in computing it separately for $q$ even or odd, and for $k = 4\ell$ and $k = 4\ell+2$.

First let us do it for the $L_k^7$ generator. Let us start with $k = 4\ell+2, q = 2l$, what one gets from \eqref{eq:polyeta} is
\begin{equation}
    \begin{aligned}
        \Delta \widetilde{\eta}^{\text{D}}_{2l}[L_{4\ell+2}^7] &= \frac{l(l-1)(4l+1)}{6} = 0 \hspace{4mm} \text{ mod 1} \,. \\
    \end{aligned}
\end{equation}
For $k = 4\ell+2, q= 2l+1$ one gets
\begin{equation}
    \begin{aligned}
        \Delta \widetilde{\eta}^{\text{D}}_{2l+1}[L_{4\ell+2}^7] &= -\frac{1+2\ell}{8}+\frac{l(l+1)(4l-1)}{6} = -\frac{1+2\ell}{8} \hspace{4mm} \text{ mod 1} \,. \\
    \end{aligned}
\end{equation}
As claimed, they only depend on the parity of $q$, not the precise value.
\\For $k = 4\ell$, $q = 2l$ one gets
\begin{equation}
    \begin{aligned}
        \Delta \widetilde{\eta}^{\text{D}}_{2l}[L_{4\ell}^7] &= -\frac{l(4l^2-1)}{6} = \frac{l}{2} = \frac{q}{4} \hspace{4mm} \text{ mod 1} \,. \\
    \end{aligned}
\end{equation}
For $k = 4\ell$, $q = 2l+1$:
\begin{equation}
    \begin{aligned}
        \Delta \widetilde{\eta}^{\text{D}}_{2l+1}[L_{4\ell}^7] &= -\frac{\ell}{4}+\frac{l(2l^2+1)}{3}  = \frac{\ell}{4} \hspace{4mm} \text{ mod 1} \,.\\
    \end{aligned}
\end{equation}
For the  $L_k^3 \times \text{K3}$ generators the story is simpler, since we find that for even $k=2\ell$ and any $q$
\begin{equation}
    \Delta \widetilde{\eta}^{\text{D}}_{q}[L_{2\ell}^3 \times \text{K3}] = 2 \Delta \widetilde{\eta}^{\text{D}}_{q}[L_{2\ell}^3] = q = 0 \hspace{4mm} \text{ mod 1} \,.
\end{equation}

\section{Constraints for higher \texorpdfstring{$k$}{k}}
\label{app:higherk}

In this appendix we extend the discussion of the main part to $k = 10$ and $12$ and summarize the anomaly constraints on the fermion spectrum.

\subsection{\texorpdfstring{$\mathbb{Z}_{10}$}{k=10}}
The bordism group splits into its coprime factors as
\begin{equation}
    \Omega_7^{\text{Spin}}(B \Z_{10}) \simeq  \Omega_7^{\text{Spin}}(B \Z_{2}) \oplus \Omega_7^{\text{Spin}}(B \Z_{5}) \simeq \Z_{16} \oplus \Z_5 \simeq \Z_{80} \,.
\end{equation}
The reduced $\eta$-invariants on the generator are
\begin{equation}
    \begin{aligned}
        \widetilde{\eta}^{\text{D}}_1 [L^7_{10} (1)] &= \widetilde{\eta}^{\text{D}}_5 [L^7_{10} (1)] = \tfrac{33}{80} \,, \quad \widetilde{\eta}^{\text{D}}_2 [L^7_{10} (1)] = \widetilde{\eta}^{\text{D}}_8 [L^7_{10} (1)] = \tfrac{1}{5} \,, \\
        \widetilde{\eta}^{\text{D}}_3 [L^7_{10} (1)] &= \widetilde{\eta}^{\text{D}}_7 [L^7_{10} (1)] =  \tfrac{1}{80} \,, \quad\widetilde{\eta}^{\text{D}}_4 [L^7_{10} (1)] = \widetilde{\eta}^{\text{D}}_6 [L^7_{10} (1)] =  \tfrac{3}{5} \, , \\
        \widetilde{\eta}^{\text{D}}_6 [L^7_{10} (1)] &=  \tfrac{13}{16} \,.
    \end{aligned}
\end{equation}
Applying the recursion relation to the fermion anomaly we extract the consistency constraint
\begin{equation}
    \mathcal{A}^F[L^7_{10} (2)] = 4 \mathcal{A}^F [L^7_{10}(1)] - \kappa j^2 \frac{12}{20} = 4 \mathcal{A}^F [L^7_{10}(1)]- \frac{3}{5}\kappa j^2 \,,
\end{equation}
which can be rewritten in terms of the fermion spectrum as
\begin{equation}
    9 (n_1 + n_3 + n_7+ n_9) + 4 (n_2 + n_4 + n _6 +n_8) + 6 n_5 = 0 \enspace \text{ mod } \, 20 \,.
\end{equation}
We can take this equation mod 4 and mod 5 to get the equivalent constraints
\begin{equation}
    \begin{aligned}
        n_1+n_3+n_5+n_7+n_9 = 0 \enspace \text{ mod } \, 4 \,, \\
        n_1+n_2+n_3+n_4+n_6+n_7+n_8+n_9 = 0 \enspace \text{ mod } \, 5 \,.
    \end{aligned}
\end{equation}
Imposing $\mathcal{A}^F[L^7_{10} (10)] = 0$ additionally leads to
\begin{equation}
    33 (n_1+n_9) +16 (n_2+n_8) + n_3+n_7 + 48(n_4+n_6) + 65n_5 = 4 + 8r \enspace \text{ mod } \, 80 \,.
\end{equation}
Taking this equation mod 8 we find
\begin{equation}
    n_1+n_3+n_5+n_7+n_9 = 4 \enspace \text{ mod } \, 8 \,,
\end{equation}
recovering the constraint under change of Spin structure discussed in Section \ref{sec:spin}.

\subsection{\texorpdfstring{$\mathbb{Z}_{12}$}{k=12}}
The bordism group splits into its coprime factors as
\begin{equation}
    \Omega_7^{\text{Spin}}(B \Z_{12}) \simeq  \Omega_7^{\text{Spin}}(B \Z_{4}) \oplus  \Omega_7^{\text{Spin}}(B \Z_{3}) \simeq \Z_{32} \oplus \Z_2 \oplus \Z_{9} \,.
\end{equation}
The reduced $\eta$-invariants are
\begin{equation}
    \begin{aligned}
        \widetilde{\eta}^{\text{D}}_1 [L^7_{12} (1)] &= \widetilde{\eta}^{\text{D}}_{11} [L^7_{12} (1)] = \widetilde{\eta}^{\text{D}}_5 [L^7_{12} (1)] = \widetilde{\eta}^{\text{D}}_{7} [L^7_{12} (1)] = \tfrac{143}{288} \,, \\
        \widetilde{\eta}^{\text{D}}_2 [L^7_{12} (1)] &= \widetilde{\eta}^{\text{D}}_{10} [L^7_{12} (1)] = \frac{19}{36}, \quad \widetilde{\eta}^{\text{D}}_3 [L^7_{12} (1)] = \widetilde{\eta}^{\text{D}}_{9} [L^7_{12} (1)] = \tfrac{23}{32} \,, \\
        \widetilde{\eta}^{\text{D}}_4 [L^7_{12} (1)] &= \widetilde{\eta}^{\text{D}}_{8} [L^7_{12} (1)] = \frac{7}{9}, \quad \widetilde{\eta}^{\text{D}}_6 [L^7_{12} (1)] = \tfrac{3}{4} \,.
    \end{aligned}
\end{equation}
The recursion relation imposes the constraint
\begin{equation}
    \mathcal{A}^F[L^7_{12} (2)] = 4 \mathcal{A}^F [L^7_{12}(1)] - \kappa j^2 \frac{12}{24} = 4 \mathcal{A}^F [L^7_{10}(1)]- \frac{j}{2}\,,
\end{equation}
which can be rewritten in terms of the fermion spectrum as
\begin{equation}
    11(n_1+n_5+n_7+n_{11})+ 8(n_2+n_4+n_8+n_{10})+ 3(n_3+n_9) = 12j \enspace \text{ mod } \, 8 \,.
\end{equation}
Taking this equation mod 8 and mod 3 leads respectively to
\begin{equation} \label{eq:recurs12}
    \begin{aligned}
        n_1+n_3+n_5+n_7+n_9+n_{11} = 4j \enspace \text{ mod } \, 8 \,, \\
        n_1+n_2+n_4+n_5+n_7+n_8+n_{10}+n_{11} = 0 \enspace \text{ mod } \, 3 \,.
    \end{aligned}
\end{equation}
On $L_{12}^3\times$K3, the relevant $\eta$-invariants read
\begin{equation}
    \begin{aligned}
        \widetilde{\eta}_1 [ L_{12}^3 \times \text{K3}] &= \widetilde{\eta}_5 [ L_{12}^3 \times \text{K3}] = \widetilde{\eta}_7 [ L_{12}^3 \times \text{K3}]= \widetilde{\eta}_{11} [ L_{12}^3 \times \text{K3}] =  \tfrac{1}{12} \\
        \widetilde{\eta}_2 [ L_{12}^3 \times \text{K3}] &= \widetilde{\eta}_4 [ L_{12}^3 \times \text{K3}] = \widetilde{\eta}_8 [ L_{12}^3 \times \text{K3}] =\widetilde{\eta}_{10} [ L_{12}^3 \times \text{K3}] = \tfrac{1}{3}  \\
        \widetilde{\eta}_3 [ L_{12}^3 \times \text{K3}] &= \widetilde{\eta}_9 [ L_{12}^3 \times \text{K3}] =  \tfrac{3}{4}
    \end{aligned}
\end{equation}
From which the fermion anomalies read:
\begin{equation}
    \begin{aligned}
        \mathcal{A}_\text{F}[ L_{12}^3(1) \times \text{K3}] = \mathcal{A}_\text{F}[ L_{12}^3(5) \times \text{K3}] &= \tfrac{1}{3}\left( n_1 +n_2 + n_4 +n_5 +n_7 + n_8 + n_{10} + n_{11}   \right) +\\ &+ \tfrac{3}{4}(n_1+n_3+n_5 + n_7 + n_9 + n_{11}) \\
        \mathcal{A}_\text{F}[ L_{12}^3(2) \times \text{K3}] = \mathcal{A}_\text{F}[ L_{12}^3(4) \times \text{K3}] &= \tfrac{1}{3}(n_1 + n_2 + n_4 + n_5 + n_7+ n_8+ n_{10} + n_{11}) \\
        \mathcal{A}_\text{F}[ L_{12}^3(3) \times \text{K3}] &= \tfrac{3}{4}(n_1 + n_3 + n_5+n_7 + n_9 + n_{11}) \\
        \mathcal{A}_\text{F}[ L_{12}^3(6) \times \text{K3}] &= 0
    \end{aligned}
\end{equation}
Which are all trivialized by the constraints \eqref{eq:recurs12}, consistently with the discussion above.

\section{Change of Spin structure}
\label{app:spinQR}

In this appendix we provide a detailed derivation of the change of the quadratic refinement under a change of Spin structure on the underlying manifold. For that we will focus on the generators of the bordism groups we encounter for the discrete gauge theories.

Lens spaces $L^n_k$ are Spin manifolds except when $n = 4m+1$ and $k$ is even \cite{FRANC1987277}. For $k$ odd the Spin structure is unique, while for $n = 4m-1$ and $k$ even there are two inequivalent Spin structures, in one-to-one correspondence with elements of $H^1(L^n_{k};\Z_2) \simeq \Z_2$.
On these elements the map \eqref{eq:bockspin} is injective. To see this consider the cohomology groups \eqref{eq:lenscoho}: for $k$ even the inclusion $\Z_2 \rightarrow\Z_k$ is non-trivial, and the Bockstein is an isomorphism since it fits into a sequence
\begin{equation}
        ... \rightarrow H^3(L_k^n;\R) \rightarrow H^3 \big(L_{k}^n;U(1)\big) \xrightarrow{\beta} H^4(L_k^n;\Z) \rightarrow H^4(L_k^n;\R) \rightarrow ...
\end{equation}
and cohomology with $\R$ coefficients is trivial on lens spaces in the relevant dimensions.

However, the particular element $\alpha^3$ we are evaluating this map on might be zero. This is the case for $k = 4 \ell$ a multiple of 4, and not for $k = 4\ell+2$. For a degree 1 co-chain $\alpha$ its Steenrod square $Sq^1(\alpha) = \alpha \cup \alpha$ is equivalent to the Bockstein homomorphism $\tilde{\beta}(\alpha)$ associated to the sequence $0\rightarrow \Z_2 \rightarrow \Z_4 \rightarrow \Z_2 \rightarrow 0$. This map sits in a long exact sequence
\begin{equation}\label{eq:bock2}
         ...\rightarrow H^1(L_{k}^n;\Z_4) \rightarrow H^1(L_{k}^n;\Z_2) \xrightarrow{\tilde{\beta}} H^2(L_{k}^n;\Z_2) \rightarrow...
\end{equation}
The universal coefficient theorem explicitly gives the cohomology in degree 1 as
\begin{equation}
    H^1(L_k^n;G) \simeq \text{Hom}\big(H_1(L_k^n ; \mathbb{Z}),G\big) \simeq \text{Hom}(\Z_k,G) \,.
\end{equation}
If $k = 4\ell$, $H^1(L_k^n;\Z_4) \simeq \Z_4$, generated by the identity map $[x]_{4\ell} \mapsto [x]_4$. This restricts mod 2 to the non-trivial element in Hom$(\Z_{4m},\Z_2) \simeq H^1(L_{k}^n;\Z_2) \simeq \Z_2$. This means that the first map in \eqref{eq:bock2} is surjective, and $\Tilde{\beta}$ is therefore zero. Thus the cup product square $\alpha \cup \alpha = 0$ and clearly so are the odd powers of $\alpha$ appearing in \eqref{eq:nushift}.

If instead $k = 4\ell+2$, $H^1(L_{k}^n;\Z_4) \simeq \Z_2$ is generated by the doubling map $[x]_{4\ell+2} \mapsto [2x]_{4}$. This is trivial when taken mod 2, so in this case $\tilde{\beta}$ is injective, and therefore $\alpha \cup \alpha$ is the non-trivial element in $H^2(L_{k}^n;\Z_2)$.

Alternatively, we can use the well known cup product structure of the $\Z_k$-valued cohomology of lens spaces (see, e.g., example 3E.2 of \cite{MR1867354}) and interpreting $\Z_2$-valued cohomology as its order 2 subgroup. $H^*(L_k^n;\Z_k)$ is generated as a ring by $x \in H^1(L_k^n;\Z_k)$ and $y = \hat{\beta}(x) \in H^1(L_k^d;\Z_k)$, with $\hat{\beta}$ the Bockstein associated to $0\rightarrow \Z \rightarrow \Z \rightarrow \Z_k \rightarrow 0$. Furthermore, $x^2 = 0$ for $k$ odd, and it is the unique order 2 element in $H^2(L_k^n;\Z_k)$ for $k$ even. 

Then, for $k$ even the element $\alpha$ associated to a change of Spin structure is $\frac{k}{2}x$, and its square is $\frac{k^2}{4}x^2$. If $k=4\ell$ the coefficient is even, and thus $\alpha^2 = 0$, while if $k = 4\ell+2$ the coefficient is odd, and $ \alpha^2 = x^2 \neq 0$.

In summary, we found that the shift in \eqref{eq:nushift} is trivial for $k=4\ell$, and an order 2 torsion class for $k = 4\ell+2$. We denote this class by $\check{b}$. 

Now let us consider a quadratic refinement associated to a differential lift $\check{\nu}$ and denote it by $\QR_\nu$. From the definition of our quadratic refinements, see also \cite{Belov:2006jd}, a shift in the differential lift amounts to \footnote{In the notation of e.g. \cite{GarciaEtxebarria:2024fuk} there is an extra factor of 2 in front of $\check{b}$ on the left hand side of \eqref{eq:qshiftb}. Since we are working with a 2-torsion characteristic classes, we omitted the formal 2 since, as can be seen from the right hand side, its contribution does not vanish even it it naively may seem so.} 
\begin{equation} \label{eq:qshiftb}
    \QR_{\nu+b}(\check{C}) = \QR_\nu(\check{C}+\check{b}) \,.
\end{equation}
Then from the defining property \eqref{eq:QRprop} we can compute the change in the associated reduced versions as
\begin{equation} 
    \widetilde{\QR}_{\nu+b}(\check{C}) = \widetilde{\QR}_{\nu}(\check{C}) + (\check{C},\check{b})
\end{equation}
The pairing is between two flat characters, so we can compute it via the torsion pairing on integral cohomology. In particular since $N_b$ is of order 2, this is
\begin{equation}\label{eq:tensSpin}
        \Delta \mathcal{A}^B_{\text{GS}} [\check{C}] = (\check{C},\check{b}) = \frac{1}{2}\int N_C \cup u
\end{equation}
where $u$ is a co-chain such that $\delta u = 2N_b$.

We can explicitly compute \eqref{eq:tensSpin} for the bordism generators by using the background $\check{C} = \check{c}*\check{c}$ lifted from a $\Z_k$ connection. On $L_k^3 \times \text{K3}$ the pairing vanishes for dimensional reasons, as argued around \eqref{eq:clens3}, while on $L_k^7$ we pick a generator and find
\begin{equation}
    \begin{aligned}
        \Delta \mathcal{A}^B_\text{GS}[L_{4\ell+2}^7 (m)] =
            \cfrac{m^2}{2}, \quad \Delta \mathcal{A}^B_\text{GS}[L_{4\ell}^7(m)] =
        \Delta \mathcal{A}^B_\text{GS}[L_k^3(m) \times \text{K3}] = 0 \,.
    \end{aligned}
\end{equation}

\end{appendix}

\bibliographystyle{JHEP.bst}
\bibliography{selfdual.bib}

\end{document}